\documentclass[12pt]{article}

  \usepackage{amsfonts,graphicx,lscape}
  \usepackage[dvips]{psfrag}

  \usepackage{amsmath}
  \usepackage{amsthm}
  \usepackage{amssymb}
  \usepackage{cite}

\newcommand{\sca}[2]{\langle #1, #2 \rangle}

\def\QK{q^{\mbox{\tiny K}}}
\def\QK{q}
\def\QC{q}

\def\PHK{\phi}
\def\PHC{\phi}

\def\CHK{\chi}
\def\CHC{\chi}

\def\RK{r}
\def\RC{r}

\newcommand{\vt}[1]{\mbox{\boldmath$#1$}}

\def\hf{\frac{1}{2}}
\def\beq{\begin{equation}} 
\def\eeq{\end{equation}} 
\def\bseq{\begin{subequations}} \def\eseq{\end{subequations}}
\def\bea{\begin{eqnarray}} \def\eea{\end{eqnarray}}
\def\bsea{\begin{subeqnarray}} \def\esea{\end{subeqnarray}}

\let\nn=\nonumber
\def\beann{\begin{eqnarray*}} \def\eeann{\end{eqnarray*}}

   \let\de=\delta
 \let\z=\zeta  
  \let\la=\lambda 
 \let\x=\xi

 \let\Ds=\displaystyle

\def\0{\over } \def\1{\vec }     \def\2{{1\over2}} \def\4{{1\over4}}
\def\5{\bar }  \def\6{\partial } \def\7#1{{#1}\llap{/}}

\def\<{\langle } \def\>{\rangle }

\def\i{{\rm i}} \def\tr{\mbox{tr}}

\def\d{{\rm d}}

\def\e{{\rm e}}

\title{
New reductions of integrable matrix PDEs \\
---$Sp(m)$-invariant systems---}

\author{
Takayuki \textsc{Tsuchida}\thanks{E-mail:\ \{surname of the author\}@ms.u-tokyo.ac.jp}
\vspace{3mm}
\\
Okayama Institute for Quantum Physics, \\
1-9-1 Kyoyama, Okayama 
700-0015, Japan}

\begin{document}
\maketitle

\begin{abstract}
We propose 
a new type of reduction for 
integrable 
systems of coupled matrix PDEs;
this reduction equates 
one matrix variable 
with the transposition of another 
multiplied by an {\it antisymmetric} \/constant 
matrix. 
Via this reduction, 
we obtain 
a new integrable 
system of coupled derivative mKdV equations and 
a new integrable 
variant 
of the massive Thirring model, 
in addition to 
the 
already known systems. 
We also 
discuss 
integrable 
semi-discretizations of 
the obtained systems 
and present 
new 
soliton solutions 
to both continuous and 
semi-discrete systems. 
As a by-product, 
a new integrable semi-discretization of the Manakov model 
(self-focusing vector NLS 
equation) is obtained.
 \end{abstract}

\vspace{1cm}
\begin{center}
\begin{minipage}{15cm}
{\sc Keywords}: \ 
$Sp(m)$-invariant systems, 
matrix derivative 
NLS 
hierarchies, 
matrix Yajima--Oikawa hierarchy, 
coupled derivative mKdV equations, massive Thirring model, 
{\it potential Kaup--Newell equation}, 
integrable discretizations, 
discrete Kaup--Newell system, 
Gel'fand--Levitan--Marchenko integral equations, 
soliton solutions
\end{minipage}
\end{center}

 \newpage
 \noindent
 \tableofcontents

 \newpage
 \noindent

\section{Introduction}
\setcounter{equation}{0}

Since the seminal work of Manakov~\cite{Manakov} in the early 70s, 
integrable 
systems of 
coupled partial differential equations (PDEs) 
that are associated with 
higher than second-order 
matrix 
spectral problems (Lax
pairs) 
have been 
the focus of intensive research. 
In particular, 
of prime importance 
among such systems 
are 
the 
vector PDEs 
that 
are 
invariant under the action of a 
classical matrix group 
on 
the vector 
dependent variables; this invariance 
also 
represents 
a ``symmetry'' 
or gauge invariance 
of the 
spectral problem. 
Because of 
this 
large 
``symmetry'', 
the vector PDEs 
usually 
allow 
the existence of solitons 
with internal degrees of freedom 
that exhibit 
highly 
nontrivial 
and interesting 
behaviors in 
soliton interactions. 
Moreover, 
their simple and symmetric form of equations 
very often 
leads to 
their 
potential or practical 
applicability in 
various branches of physics as well as 
applied mathematics. 
Typical 
examples of such vector PDEs 
include 
the 
$U(m)$-invariant 
Manakov model~\cite{Manakov,ZaSha}, 
also referred to as 
the self-focusing vector nonlinear Schr\"{o}dinger (NLS) 
equation, 
\begin{equation}
 {\rm i} \vt{q}_{t} + \vt{q}_{xx} 
  + 2 \| \vt{q} \|^2 
\vt{q} = \vt{0},
\hspace{5mm} 
\vt{q} = (q_1, q_2, \ldots, q_m), 
\hspace{5mm} \|\vt{q}\|^2 
:=\vt{q} \cdot \vt{q}^\dagger = \sum_{j=1}^m |q_j|^2;
\label{cNLS}
\end{equation}
two distinct 
versions of the 
vector mKdV equation~\cite{Svip,YO2,Konop3}
with $O(m)$-invariance, 
\begin{align}
& \vt{q}_{t} + \vt{q}_{xxx} 
  + 6 \sca{\vt{q}}{\vt{q}}\vt{q}_x = \vt{0},
\hspace{5mm} 
\vt{q} = (q_1, q_2, \ldots, q_m), 
\hspace{5mm} 
\sca{\vt{q}}{\vt{q}} 
:= \vt{q} \cdot \vt{q}^T = \sum_{j=1}^m q_j^2, 
\label{vmKdV1}
\\
& \vt{q}_{t} + \vt{q}_{xxx} 
  + 3 \sca{\vt{q}}{\vt{q}_x}\vt{q}
  + 3 \sca{\vt{q}}{\vt{q}}\vt{q}_x 
= \vt{0};
\nonumber
\end{align}
the vector 
third-order 
Heisenberg ferromagnet model 
with $O(m)$-invariance~\cite{Adler,Tsuchida5}, 
\begin{equation}
\vt{S}_t + \vt{S}_{xxx} + \frac{3}{2} (\sca{\vt{S}_x}{\vt{S}_x} \vt{S} )_x 
	= \vt{0}, 
\quad 
\sca{\vt{S}}{\vt{S}} 
=1;
\label{hoHF}
\end{equation}
and 
the 
$O(m)$-invariant vector 
extension~\cite{Tsuchida5} 
of the third-order 
Wadati--Konno--Ichikawa
equation~\cite{WKI},
\[
\vt{q}_t + 
\left[ \frac{\vt{q}_x}{(1-\sca{\vt{q}}{\vt{q}})^{\frac{3}{2}}} 
\right]_{xx} = \vt{0}.
\]
The recent developments 
in computer algebra packages and 
improvements in 
CPU 
performances 
have 
further 
increased the number 
of $m$-component 
integrable systems with $U(m)$- 
or $O(m)$-invariance 
as well as 
\mbox{$(m+1)$}-component systems 
with $O(m)$-invariance to a considerable extent~\cite{SW,TsuWo,AnWo}. 
However, 
in contrast to the 
$U(m)$-invariant 
and 
$O(m)$-invariant systems, 
little research 
has been conducted 
on 
integrable systems 
having invariance with respect to 
the symplectic group $Sp(m)$; 
this 
refers to 
the group of \mbox{$2m \times 2m$} real/complex 
symplectic matrices, 
$Sp(m,{\mathbb R})$ or $Sp(m,{\mathbb C})$, 
in accordance with the 
attribute of the dependent variables. 
To the best of the author's knowledge, 
the {\it only} \/example
of 
a $2m$-component 
vector nonlinear 
PDE 
with $Sp(m)$ invariance 
is the system of coupled derivative 
mKdV equations studied using the 
bilinear 
method by Iwao and Hirota~\cite{Iwao2},
\begin{equation}
\frac{\6 u_{i}}{\6 t} + \frac{\6^3 u_{i}}{\6 x^3} 
 + 3 \Biggl[ \sum_{j=1}^m
	\Bigl( \frac{\6 u_{2j-1}}{\6 x} u_{2j} - u_{2j-1} 
	\frac{\6 u_{2j}}{\6 x} \Bigr) \Biggr] \frac{\6 u_{i}}{\6 x} = 0,
\hspace{5mm} 
  i=1, 2, \ldots, 2m,
\label{cdmkdv01}
\end{equation}
as well as its higher symmetries. 
It should be noted that 
system (\ref{cdmkdv01}) is a natural multi-component 
generalization of 
the two-component system [(\ref{cdmkdv01}) with $m=1$] 
derived within the framework of the 
Sato theory 
by Loris and Willox~\cite{Loris1,Loris3}. 
Here, the $Sp(m)$ 
invariance 
refers to the fact that 
system (\ref{cdmkdv01}) 
is form-invariant under the 
following 
linear 
transformation:
\mbox{$(u_1, u_2, \ldots, u_{2m})\mapsto (u_1, u_2, \ldots, u_{2m}) 
\hspace{1pt}S^T$}, 
where $S$ 
is an 
\mbox{$(x,t)$}-independent element of 
the symplectic group \mbox{$Sp(m)$} 
with 
a proper 
ordering 
of the 
base vectors, 
and the superscript ${}^T$ denotes the transposition. 

The main objective 
of this paper is to expand 
the class of 
$Sp(m)$-invariant 
integrable systems 
and to characterize 
some of their interesting 
properties 
within the framework of the inverse scattering method. 
To achieve this goal, 
it must 
be first noted 
that system (\ref{cdmkdv01}) is homogeneous 
with respect to 
the following 
weighting scheme: 
\mbox{$w(\6_x) =1$}, \mbox{$w(\6_t) =3$}, and 
\mbox{$w(u_i) =1/2$}; 
this 
weighting scheme 
is the same 
as that for the 
third-order symmetries of 
vector derivative NLS (DNLS)-type systems. 
Motivated by this observation, 
we start with 
more general systems, 
that is, the 
third-order symmetries of 
integrable 
matrix generalizations 
of the 
DNLS-type equations. 
Then, we 
propose a {\it new} \/type 
of reduction for 
these 
third-order 
integrable matrix PDEs, 
at least in its explicit form, 
wherein 
one matrix variable 
is related to the 
transposition of the other 
multiplied by an antisymmetric constant matrix~\cite{Gakkai}. 
Considering the special case 
wherein the matrix 
variables are restricted to the form of 
column/row vectors, we 
obtain 
two $Sp(m)$-invariant 
systems; 
one coincides with 
system (\ref{cdmkdv01}), while 
the other one, 
\begin{equation}
\frac{\6 u_{i}}{\6 t} + \frac{\6^3 u_{i}}{\6 x^3} 
 + 3 \frac{\6}{\6 x} 
\Biggl[ \sum_{j=1}^m
	\Bigl( \frac{\6 u_{2j-1}}{\6 x} u_{2j} - u_{2j-1} 
	\frac{\6 u_{2j}}{\6 x} \Bigr) u_{i} \Biggr] 
	= 0,
\hspace{5mm} 
 i=1, 2, \ldots, 2m,
\label{cdmkdv02}
\end{equation}
appears 
to be a new integrable system~\cite{Gakkai}. 
Note 
that the location 
of $\6/\6 x$ in 
(\ref{cdmkdv02}) 
is different from that in 
(\ref{cdmkdv01}). 

Once we have identified systems (\ref{cdmkdv01}) and (\ref{cdmkdv02}) 
as the 
reductions of the 
matrix DNLS-type systems, it is not 
difficult to further extend 
the class of $Sp(m)$-invariant 
integrable systems. 
First, we consider the 
integrable matrix generalizations~\cite{Nij1,Tsuchida3} 
of massive Thirring-type models~\cite{Kuz,KN2,KaMoIno,GIK,NCQL}, which are 
the first negative flows 
of the matrix DNLS
hierarchies. 
Then, via the same type of 
reduction, we 
can directly 
obtain 
a new integrable variant of the massive Thirring model that 
is a hyperbolic 
system 
with $Sp(m)$-invariance, 
\begin{equation}
\frac{\6^2 v_{i}}{\6 \tau \6 x} 
+ 
v_{i} - \Biggl[ \sum_{j=1}^m 
\Bigl( \frac{\6 v_{2j-1}}{\6 x} v_{2j} - v_{2j-1} \frac{\6 v_{2j}}{\6 x}\Bigr) 
 \Biggr] v_{i} = 0,
\hspace{5mm} 
i=1, 2, \ldots, 2m,
\label{cdMTM0}
\end{equation}
as well as an equivalent system 
up to the interchange of 
$x$ 
and 
$\tau$. 
Second, we 
consider the 
integrable space discretizations 
(semi-discretizations, for short) of the matrix 
DNLS hierarchies, including the massive Thirring-type models, 
proposed 
in ref.~\citen{Tsuchida4}. 
Though 
not all of the semi-discrete matrix DNLS-type systems 
in ref.~\citen{Tsuchida4} are 
useful for our purpose, 
we find that the semi-discrete 
Kaup--Newell hierarchy 
[cf.\ (3.1) and (5.2) 
in ref.~\citen{Tsuchida4}] 
allows 
proper reductions to 
yield integrable 
semi-discretizations of 
the 
above 
$Sp(m)$-invariant 
systems. 
Note that the discrete analogue of the 
reduction used to obtain 
a continuous $Sp(m)$-invariant system 
may 
not 
be uniquely determined. 
In fact, 
in addition to 
a single integrable 
semi-discretization of 
system (\ref{cdmkdv01}), 
we 
obtain {\it two} \/integrable 
semi-discretizations for 
each of the $Sp(m)$-invariant 
systems (\ref{cdmkdv02}) and (\ref{cdMTM0}). 
This result partly 
illustrates 
the 
wide applicability of the 
semi-discrete 
Kaup--Newell 
hierarchy 
provided in ref.~\citen{Tsuchida4}
to the theory of 
integrable discretizations. 

Hereafter, 
we 
will not restrict ourselves 
to 
the {\em canonical} \/representation 
of the 
$Sp(m)$-invariant systems, 
but will present 
them 
in a slightly 
generalized (but still being 
integrable) form. 
Specifically, we 
replace 
terms 
such as 
\mbox{$\sum_{j=1}^m 
\bigl(
\frac{\6 u_{2j-1}}{\6 x} u_{2j} - u_{2j-1} \frac{\6 u_{2j}}{\6 x} 
\bigr)$} by 
those such as 
\mbox{$\sum_{1 \le j<k \le M} C_{jk} 
\bigl( 
\frac{\6 u_{j}}{\6 x} u_{k} - u_{j} \frac{\6 u_{k}}{\6 x} 
\bigr)$}, 
where 
the integer 
$M$ may or may not be even, and 
$C_{jk}$ (\mbox{$j<k$}) are arbitrary 
coupling constants (cf.\ ref.~\citen{Iwao2}). 
It can be recalled 
that
according to 
the classical 
theory of matrices 
(see, {\em e.g.}, refs.~\citen{Satake,Serre}), 
any 
antisymmetric 
matrix 
\mbox{$C :=(C_{jk})_{j,k=1, \ldots ,M}$}
can be 
transformed 
to the block diagonal 
form 
\[
P^T C P = 
\hspace{4.5mm}
\overbrace{ 
\hspace{-4.5mm}
\left(
\begin{array}{cccc}
J &  & & \\
 &  \ddots &  & \\
& & J & \\
& &  &  
\mbox{\Large $O$} \\
\end{array}
\right), 
\hspace{-14.5mm}}^{m}\hspace{14.5mm}
\quad 
J =
\left(
\begin{array}{cc}
0 & 1\\
-1 & 0 \\
\end{array}
\right),
\]
with an invertible matrix $P$. 
Here, 
\mbox{$2m \,(\le \hspace{-1pt}M)$} is equal to 
the rank of 
$C$, 
where $m$ denotes the number of copies of $J$. 
This guarantees that 
through an invertible 
linear change 
of 
the dependent variables, 
the generalized system 
involving 
$C_{jk}$ 
can always be converted 
into an 
$Sp(m)$-invariant (sub)system in 
the 
canonical form with 
linear equation(s) 
coupled to it, if any (the case of \mbox{$2m\hspace{-1pt} < \hspace{-1pt}M$}). 
Therefore, 
in this paper, 
we also 
use the term ``symplectic invariance'' 
for such generalized systems. 

Using an approach based on the inverse scattering method, 
we can construct 
multi-soliton solutions of 
the matrix DNLS hierarchies 
under appropriate boundary conditions 
in both 
the continuous and 
discrete cases~\cite{Tsuchi05}. 
Thus, the 
soliton solutions 
for 
the $Sp(m)$-invariant systems 
contained in 
these 
hierarchies 
can be obtained 
by 
properly specializing the 
soliton 
parameters 
involved 
in the solutions 
for the latter. 
Moreover, as one of the main advantages of the inverse scattering method, 
the {\it most general} \/soliton solutions 
are obtained under the specified boundary conditions. 
However, the computations and discussions 
required 
to 
arrive at 
simple 
explicit 
formulas for 
the multi-soliton solutions 
are rather 
extensive 
and involved; 
hence, 
we do not 
present them here. 
We 
will defer 
the detailed derivation and investigation of 
the multi-soliton solutions 
to a subsequent 
publication~\cite{Tsuchi05}. 
In this paper, 
we assume decaying boundary conditions 
at spatial 
infinity and 
start with the 
linear integral/summation 
equations 
associated with the continuous/discrete 
matrix DNLS hierarchies, 
omitting 
their derivation 
via the inverse scattering method. 
These linear integral/summation equations, 
referred to as 
the Gel'fand--Levitan--Marchenko type, 
provide an 
{\it exact linearization}~\cite{ARS} 
of 
each 
nonlinear system under 
study 
in that 
they 
provide a
relation between 
the solutions of 
the 
nonlinear system 
and those 
of the corresponding 
linear system. 
Considering a 
special 
case 
with 
a proper reduction, 
we 
solve 
the 
Gel'fand--Levitan--Marchenko 
equations 
to obtain 
the one-soliton solutions 
of the $Sp(m)$-invariant systems 
under the 
decaying boundary conditions. 
It should be emphasized 
that the 
soliton solutions 
of (\ref{cdmkdv01}) 
obtained in this manner 
are 
indeed {\em more general} \/than 
the 
previously known 
solutions~\cite{Loris1,Iwao2}. 
Although the accuracy 
of these 
one-soliton solutions 
can easily be 
verified by direct substitutions, 
unlike the case of the NLS equation, 
it is not 
easy to 
obtain 
such solutions directly 
without 
resorting to the inverse scattering method 
or other 
sophisticated 
methods in soliton theory. 
Indeed, any 
{\it naive} 
\/ansatz for the 
travelling 
wave solutions, 
{\it e.g.}, a 
complex 
plane wave 
modulated by a real 
envelope 
moving with a 
constant velocity, 
is most likely to 
fall into a trivial 
subclass of the 
most 
general one-soliton solutions, 
such as ``the one-soliton solution'' 
proposed 
by Loris and Willox~\cite{Loris1}. 

This paper is organized as follows. 
In section 2, we demonstrate 
that the continuous 
systems 
(\ref{cdmkdv01}), (\ref{cdmkdv02}), and (\ref{cdMTM0}) 
in a slightly generalized form, 
as mentioned above, 
are 
obtained 
through a new type of reduction of 
the 
matrix DNLS hierarchies. 
We also present their 
bright 
one-soliton solutions. 
In section 3, we propose 
the 
integrable semi-discretizations of 
these 
continuous systems 
and present the 
one-soliton solutions 
for 
the most 
interesting 
semi-discrete systems. 
We also 
derive 
a new integrable semi-discretization of the Manakov model (\ref{cNLS}) 
from one of these semi-discrete systems. 
The last section, section 4, is devoted to concluding remarks, 
wherein we 
state that 
the 
reduction considered in this paper is not 
restricted to the 
DNLS-type systems, but 
is also applicable to 
a matrix generalization of the Yajima--Oikawa hierarchy. 

\section{Reductions of continuous 
matrix DNLS 
hierarchies}
\label{}
\setcounter{equation}{0}
In this section, we propose the 
$Sp(m)$-invariant integrable systems 
via reductions of the 
continuous matrix 
derivative NLS hierarchies. 
We also present the 
Lax pairs, associated linear integral equations, 
and 
one-soliton solutions for the obtained 
systems.

\subsection{Coupled derivative 
mKdV equations 
of 
Chen--Lee--Liu type}
\label{CLLtype}

\subsubsection{Lax pair for the 
second flow 
of the matrix 
Chen--Lee--Liu hierarchy}

A derivative nonlinear Schr\"{o}dinger (DNLS) equation 
\mbox{$\i \QC_{t_2} + \QC_{xx} \pm \i |\QC|^2 \QC_x = 0$}, 
which is often 
referred to as 
the Chen--Lee--Liu equation~\cite{CLL}, permits 
an 
integrable matrix generalization~\cite{Nij1,Olver,Tsuchida3,Dimakis} 
\begin{equation}
\left\{ 
\hspace{-1pt}
\begin{array}{l}
\i \QC_{t_2} + \QC_{xx} -\i \QC \RC \QC_x = O, \\[1pt]
\i \RC_{t_2} - \RC_{xx} -\i \RC_x \QC \RC = O.
\end{array}
\right.
\label{mCLL}
\end{equation}
Here, $\QC$ and $\RC$ are \mbox{$l_1 \times l_2$} and \mbox{$l_2 \times l_1$} 
matrices, respectively. 
Note that 
$O$ 
on the right-hand side
of the 
equations 
implies that the dependent variables can take 
their 
values in 
matrices. 
The lower indices of $t$ 
are used to 
distinguish between 
different (and commutative) 
time evolutions; however, 
in the following text, we 
often omit these indices for 
brevity. 
In this paper, we are more concerned 
with the next higher flow in the matrix Chen--Lee--Liu hierarchy 
that commutes with the first 
nontrivial 
flow (\ref{mCLL}). 
It is written as 
(up to a scaling of the time variable)~\cite{Nij1,Olver,Dimakis} 
\begin{equation}
\left\{ 
\hspace{1pt}
\begin{split}
& \QC_{t_3} + \QC_{xxx} - \i \frac{3}{2} (\QC_x \RC \QC_x + \QC \RC \QC_{xx})
 -\frac{3}{4} \QC \RC \QC \RC \QC_x = O, \\[2pt]
& \RC_{t_3}
 + \RC_{xxx} + \i \frac{3}{2} (\RC_x \QC \RC_x + \RC_{xx} \QC \RC)
 -\frac{3}{4} \RC_x \QC \RC \QC \RC = O.
\end{split}
\right.
\label{higherCLL}
\end{equation}
The Lax pair 
for (\ref{higherCLL}) is given by 
\bseq
\begin{align}
U = \mbox{}& 
\i \z^2 \left[
\begin{array}{cc}
 -I_1  &  \\
    &  I_2 \\
\end{array}
\right]
+ 
\z \left[
\begin{array}{cc}
   &  \QC \\
 \RC  &   \\
\end{array}
\right]
+ \i
\left[
\begin{array}{cc}
  O &  \\
    & \frac{1}{2}\RC \QC  \\
\end{array}
\right],
\label{U_form}
\\[8pt]
V = \mbox{}&
\i \z^6 
\left[
\begin{array}{cc}
 -4I_1 &  \\
   & 4I_2  \\
\end{array}
\right]
+\z^5
\left[
\begin{array}{cc}
  & 4\QC \\
 4\RC &  \\
\end{array}
\right]
+\i \z^4
\left[
\begin{array}{cc}
-2\QC\RC  &  \\
  & 2\RC\QC \\
\end{array}
\right]
+ \z^3
\left[
\begin{array}{cc}
  & 2\i \QC_x + \QC \RC \QC \\
 -2\i \RC_x + \RC \QC \RC &  \\
\end{array}
\right]
\nn \\
&
\mbox{}
+\i \z^2
\left[
\begin{array}{cc}
 -\i (\QC_x \RC - \QC \RC_x) -\hf (\QC \RC)^2 &  \\
  & \i (\RC \QC_x - \RC_x \QC) + \hf (\RC \QC)^2 \\
\end{array}
\right]
\nn \\
& \mbox{} + \z 
\left[
\begin{array}{cc}
 & -\QC_{xx} + \frac{\i}{2} (\QC_x \RC \QC - \QC \RC_x \QC +2\QC \RC \QC_x) 
+ \frac{1}{4} (\QC\RC)^2 \QC \\ 
-\RC_{xx} - \frac{\i}{2} (2\RC_x \QC \RC - \RC \QC_x \RC + \RC \QC \RC_x) 
+ \frac{1}{4} (\RC\QC)^2 \RC &
\end{array}
\right]
\nn \\
& \mbox{} + \i
\left[
\begin{array}{cc}
O & \\
 & - \frac{1}{2} (\RC_{xx} \QC -\RC_x \QC_x + \RC \QC_{xx})
 - \frac{\i}{4} (2 \RC_x \QC \RC \QC -\RC \QC_x \RC \QC + \RC \QC \RC_x \QC 
 - 2 \RC \QC \RC \QC_x) + \frac{1}{8} (\RC \QC)^3
\end{array}
\right].
\label{V_form2}
\end{align}
\label{LaxCLL}
\eseq
Here, $\z$ is 
the 
spectral parameter 
independent of $x$ and $t$; 
$I_1$ and $I_2$ are 
the \mbox{$l_1 \times l_1$} and 
\mbox{$l_2 \times l_2$} unit 
matrices, respectively. 
Let us 
substitute the Lax pair (\ref{LaxCLL}) 
in
the zero-curvature condition~\cite{AKNS74,ZS79}
\beq
U_t -V_x +UV-VU = O, 
\label{Lax_eq}
\end{equation}
which is the compatibility condition 
for the overdetermined 
system of linear 
PDEs, 
\begin{equation}
\Psi_x = U \Psi, \hspace{5mm} \Psi_t = V \Psi.
\label{lin}
\end{equation}
Subsequently, 
equating the terms with the same powers of $\z$ to zero, we 
obtain the third-order 
matrix Chen--Lee--Liu 
system (\ref{higherCLL}) without any contradiction or additional 
constraint. 

\subsubsection{Reduction}
\label{rCLL}
Both the first flow 
(\ref{mCLL}) and 
the second flow 
(\ref{higherCLL}) 
of the matrix Chen--Lee--Liu hierarchy 
permit the reduction of the 
Hermitian conjugation~\cite{Makhankov} 
\mbox{$\RC = A_1 \QC^\dagger A_2$}, where $A_1$ and $A_2$ 
are constant Hermitian matrices:\ 
{\mbox{$A_i^{\dagger}=A_i$} and 
\mbox{$A_{i,t}=A_{i,x}=O$}. 
Moreover, 
unlike the first flow (\ref{mCLL}), 
the second flow 
(\ref{higherCLL}) 
allows an 
interesting 
reduction such 
that $\RC$ is identically 
equal to 
$C \QC^T$, where $C$ is 
an antisymmetric constant matrix:\ 
{\mbox{$C^T=-C$}, \hspace{1pt}\mbox{$C_t=C_x=O$}. 
In fact, 
the reduction 
\mbox{$\RC 
=\QC^T C$} 
(or, more generally, \mbox{$\RC = B \QC^T C$}, \hspace{1pt}
\mbox{$B^T=B$}, \hspace{1pt}\mbox{$C^T=-C$})
can also be considered 
for (\ref{higherCLL}), 
but we do not exploit
this reduction 
in order to 
maintain a 
natural and easy-to-read
flow of the paper. 
For the reduced system to 
assume a concise form without the imaginary unit and fractions, 
we consider the following vector reduction 
($l_1=1$, $l_2=M$): 
\begin{equation}
\QC = (u_{1}, \ldots, u_{M}), \hspace{5mm} 
\RC = 2\i C \QC^T 
= 2\i 
\left(
\begin{array}{c}
 \sum_{k=1}^M C_{1 k} u_{k}  \\
  \vdots \\
 \sum_{k=1}^M C_{M k} u_{k}  \\
\end{array}
\right),
\label{setting}
\end{equation}
which also implies 
the simple relation \mbox{$\QC \RC =0$}. 
System (\ref{higherCLL}) is 
then reduced to a system of 
coupled derivative mKdV equations~\cite{Iwao2}:
\begin{equation}
\frac{\6 u_{i}}{\6 t} + \frac{\6^3 u_{i}}{\6 x^3} 
 + 3 \Biggl[ \sum_{1 \le j<k \le M} 
	C_{jk} \Bigl( \frac{\6 u_{j}}{\6 x} u_{k} - u_{j} 
	\frac{\6 u_{k}}{\6 x} \Bigr) \Biggr] \frac{\6 u_{i}}{\6 x} = 0,
\hspace{5mm} 
i=1, 2, \ldots, M.
\label{cdmkdv1}
\end{equation}
Using the vector notation, (\ref{cdmkdv1}) can also be written 
as 
\[
\vt{u}_t + \vt{u}_{xxx} + 3\sca{\vt{u}_x C}{\vt{u}} \vt{u}_x 
= \vt{0}, \hspace{5mm} C^T = -C. 
\]
Here, \mbox{$\vt{u}=(u_{1}, u_{2}, \ldots, u_{M})$} and 
\mbox{$\sca{\vt{u}_x C}{\vt{u}} = \vt{u}_x C \vt{u}^T$}. 
The Lax pair for 
this 
reduced system 
is 
obtained by substituting (\ref{setting}) into 
$\QC$ and $\RC$ in (\ref{LaxCLL});
using 
a 
simple 
gauge transformation, 
the Lax pair 
can be
rewritten in the form
\bseq
\begin{align}
U = \mbox{}& 
\left[
\begin{array}{cc}
 -\lambda &  \vt{u} \\
 \lambda C\vt{u}^T  &  - C \vt{u}^T \vt{u} \\
\end{array}
\right],
\label{spec1}
\\[8pt]
V = \mbox{}&
\left[
\begin{array}{c|c}
 \lambda^3 + 2\lambda \sca{\vt{u}_x C}{\vt{u}}_{\vphantom \int_{\vphantom 0}} 
& -\lambda^2\vt{u} +\lambda \vt{u}_x - \vt{u}_{xx} 
 -2 \sca{\vt{u}_x C}{\vt{u}}_{\vphantom \int_{\vphantom 0}} \vt{u}  \\
\hline
\begin{array}{l}
\, -\lambda^3 C\vt{u}^T -\lambda^2 C\vt{u}_x ^T 
 \vphantom{C^{\vphantom{\int}^{\vphantom{\int}}}}
\\ \mbox{} 
-\lambda C\vt{u}^T_{xx} 
-2\lambda \sca{\vt{u}_x C}{\vt{u}} C \vt{u}^T 
\end{array}
 & 
\begin{array}{l}
 \, \lambda^2 C\vt{u}^T \vt{u} 
	+ \lambda C (\vt{u}^T_x \vt{u} - \vt{u}^T \vt{u}_x) 
 \vphantom{C^{\vphantom{\int}^{\vphantom{\int}}}}
\\ \mbox{} 
 + C (\vt{u}^T_{xx} \vt{u} -\vt{u}^T_{x} \vt{u}_{x}
 + \vt{u}^T \vt{u}_{xx})
 +2 \sca{\vt{u}_x C}{\vt{u}} C\vt{u}^T \vt{u} \\
\end{array}
\end{array}
\right],
\end{align}
\label{LaxCLL2}
\eseq
where \mbox{$\lambda:=2\i \z^2 $} is the ``new'' 
spectral parameter. 
In the simplest nontrivial case of 
\mbox{$M=2$}, 
setting 
\[
C_{12}= \i, \hspace{5mm} 
u_1 = \psi, \hspace{5mm} u_2 = \psi^\ast,
\]
we obtain 
the following single equation~\cite{Loris1,Loris3}:
\begin{equation}
\psi_t + \psi_{xxx} + 3 \i (\psi_x \psi^\ast - \psi \psi^\ast_x) \psi_x = 0.
\label{single1}
\end{equation}
Here, the asterisk denotes the 
complex conjugate. 
Using a 
simple point transformation (cf.\ refs.~\citen{IKWS,Kawata2}), 
we can 
convert (\ref{single1}) into 
an NLS-type equation 
perturbed by higher order terms. 

\subsubsection{Relation to the second flow of the 
matrix NLS hierarchy}

The matrix 
DNLS hierarchies, including the matrix Chen--Lee--Liu hierarchy, 
can be considered 
embedded in a generalization 
of the matrix NLS hierarchy~\cite{Tsuchida4}. 
In the case of reduction 
(\ref{setting}), 
the particular 
relation $qr=0$ 
simplifies the embedding formulas to a considerable 
extent. 
As a result, 
(\ref{cdmkdv1}) can be derived from 
the second nontrivial flow of the matrix NLS hierarchy, i.e., 
a matrix analogue 
of the (non-reduced) complex mKdV equation~\cite{Zakh,Konop1}
\begin{equation}
\left\{ \hspace{-1mm}
\begin{array}{l}
 Q_t + Q_{xxx} -3 Q_x RQ - 3 QR Q_x = O, \\
 R_t + R_{xxx} -3 R_x QR - 3 RQ R_x = O, 
\end{array}
\right.
\label{cmkdv}
\end{equation}
through a simple, 
but not ultralocal, reduction. 
Indeed, 
if we set
\begin{equation}
Q = (u_{1}, \ldots, u_{M}), \hspace{5mm} 
R = C Q_x^T
= \left(
\begin{array}{c}
 \sum_{k=1}^M C_{1 k} u_{k,x}  \\
  \vdots \\
 \sum_{k=1}^M C_{M k} u_{k,x}  \\
\end{array}
\right),
\label{red1}
\end{equation}
in terms of an antisymmetric constant matrix $C$, 
we obtain 
the following relations:
\[
Q_x R =0, \hspace{5mm} 
Q R = - \sum_{1 \le j < k \le M} C_{jk} (u_{j,x} u_k 
	- u_j u_{k,x}), 
\hspace{5mm}
Q R_x = (QR)_x. 
\]
Thus, 
the two matrix equations (\ref{cmkdv}) simply 
collapse to form 
a single vector equation (\ref{cdmkdv1}). 

There exist a few 
advantages in regarding (\ref{cdmkdv1}) as a 
reduced form 
of (\ref{cmkdv}). 
First, 
an infinite set of conservation laws for (\ref{cmkdv}) 
can be constructed systematically and rather easily 
using a 
recursive 
formula based on the Lax pair~\cite{Tsuchida1}. 
Thus, we can obtain the 
conservation laws for (\ref{cdmkdv1}) 
from those for (\ref{cmkdv}) 
through the reduction (\ref{red1}). 
The conserved densities of the first two ranks obtained in 
this manner 
are given by 
\[
u_{j,x} u_{k} \;\, (j  \neq k), \hspace{5mm} 
\sum_{1 \le j < k \le M} C_{jk} (u_{j,xx} u_{k,x} 
	- u_{j,x} u_{k,xx})
-\left[ \sum_{1 \le j < k \le M} C_{jk} (u_{j,x} u_k 
	- u_j u_{k,x}) 
\right]^2, 
\]
which are derived 
from the conserved densities 
\mbox{$RQ$} and \mbox{$\tr \hspace{1pt} ( Q_x R_x + QRQR )$} 
of (\ref{cmkdv}), 
respectively. 
Second, 
considering 
a similar 
reduction for 
a space-discrete 
analogue of (\ref{cmkdv})~\cite{Tsuchida4}, 
we can obtain an integrable semi-discretization 
of the continuous 
system (\ref{cdmkdv1}). 
We employ this approach in subsection~\ref{latCLL} 
since none of the 
(semi-)discrete analogues of the 
third-order matrix Chen--Lee--Liu 
system (\ref{higherCLL}), which 
are integrable and 
permit a 
reduction like (\ref{setting}), 
are known. 

\subsubsection{Solution formulas}
An approach 
based on the 
inverse scattering method~\cite{Tsuchi05} enables 
the derivation of 
the solutions of 
the matrix 
Chen--Lee--Liu hierarchy, that is, 
(\ref{mCLL}) and 
(\ref{higherCLL}) 
as well as 
the higher 
flows, 
through a 
set of formulas (cf.\ ref.~\citen{Kawa} 
for the case of scalar variables) 
\begin{subequations}
\begin{align}
q
&=
K(x, x),
\\[2pt]
r
&=
\bar{K}(x, x),
\\[2pt]
K(x, y) 
&=
 \bar{F} (y)
 + \frac{\i}{2} 
\int^\infty_x \d s_1 \int^\infty_x \d s_2\hspace{1pt} 
K (x, s_1)
F (s_1+s_2-x) \frac{\6 \bar{F} (s_2+y-x)}{\6 s_{2}},
\hspace{5mm} y \ge x,
\label{Gel1c}
\\
\bar{K}(x, y) 
&=
 F (y)
 - \frac{\i}{2} \int^\infty_x \d s_1 \int^\infty_x \d s_2\hspace{1pt} 
\frac{\6 \bar{K} (x, s_1)}{\6 s_1} 
\bar{F}(s_1+s_2-x) F (s_2+y-x),
\hspace{5mm} y \ge x.
\label{Gel1d}
\end{align}
\label{Gel1}
\end{subequations}
Here and hereafter, the bar does {\it not} \/denote the 
complex/Hermitian conjugate in general. 
The time dependence of the 
functions 
is suppressed 
in 
formulas 
(\ref{Gel1}). 
The 
functions 
$\bar{F}(x
)$ and $F (x
)$ 
satisfy the corresponding 
{\it linear} \/uncoupled 
system of 
matrix PDEs, 
{\em e.g.}, 
\begin{equation}
\i \frac{\6 \bar{F}}{\6 t_2} + \frac{\6^2 \bar{F}}{\6 x^2}
=O, 
\hspace{5mm}
\i \frac{\6 F}{\6 t_2} - \frac{\6^2 F}{\6 x^2}=O
\label{lin1}
\end{equation}
for the second-order matrix Chen--Lee--Liu system (\ref{mCLL}) 
and 
\begin{equation}
\frac{\6 \bar{F}
}{\6 t_3} + \frac{\6^3 \bar{F}
}{\6 x^3}=O, 
\hspace{5mm}
\frac{\6 F
}{\6 t_3} + \frac{\6^3 F
}{\6 x^3}=O
\label{lin2}
\end{equation}
for the third-order matrix Chen--Lee--Liu flow (\ref{higherCLL}), 
and decay rapidly as \mbox{$x \to + \infty$}. 
Note that 
formulas (\ref{Gel1}) 
involve only 
equal-time 
quantities 
and 
depend on the time variables 
through 
the 
time evolution
of $\bar{F}$ and $F$. 

The reduction 
(\ref{setting}) is achieved 
at the level of 
the 
solution formulas 
by 
setting 
\[
\bar{F}(x,t) = (f_1, f_2, \ldots, f_M)(x,t)
=: \vt{f}(x,t), 
\hspace{5mm}
F(x,t) = 2\i C \bar{F}(x,t)^T.
\]
With this reduction, the 
set of 
formulas (\ref{Gel1}) is 
reduced to a compact form. 
Thus, 
the solutions 
to the coupled derivative mKdV equations (\ref{cdmkdv1}), 
decaying 
as \mbox{$x \to + \infty$}, 
can be constructed 
from 
those of the 
linear vector PDE 
\mbox{$\vt{f}_t + \vt{f}_{xxx}=\vt{0}$} 
through the formula
\begin{subequations}
\begin{align}
\vt{u}(x,t)
&=
\vt{k}(x, x
;t
),
\\[2pt]
\vt{k}(x, y) 
&=
 \vt{f} (y) -
\int^\infty_x \d s_1 \int^\infty_x \d s_2\hspace{1pt}
\vt{k}(x, s_1) C \vt{f} (s_1+s_2-x)^T
\frac{\partial \vt{f} (s_2+y-x)
}{\partial s_2}
, \hspace{5mm} y \ge x.\hspace{5mm}
\label{Gel2}
\end{align}
\label{Gelf}
\end{subequations}
Here, \mbox{$\vt{u}
= (u_1, u_2, \ldots, u_M)$} and 
$\vt{k}(x,y)
$ are $M$-component row vectors. 
Note that 
for a given $\vt{f}$, 
the integral equation 
(\ref{Gel2}), 
referred to as 
the Gel'fand--Levitan--Marchenko type, 
is linear for 
unknown $\vt{k}$. 
We 
assume that 
both $\vt{u}$ and 
the 
eigenfunctions 
bound 
in the potential $\vt{u}$ 
(cf.\ (\ref{lin})) 
should also 
decay 
as \mbox{$x \to - \infty$}. 
Then, 
we can verify 
that the 
number of 
distinct 
exponential 
functions 
comprising 
$\vt{f}(x,t)$ 
has to 
be even. 
Substituting 
\begin{align}
\vt{f} (x,t) &= \vt{a}_1
\e^{\i \la_1 x + \i \la_1^3 t} + \vt{a}_2 
\e^{\i \la_2 x + \i \la_2^3 t}, 
\quad
{\rm Im}\, \la_j>0 \;(j=1,2), 
\quad
\la_1 \neq \la_2, 
\quad 
\langle \vt{a}_1 C, \vt{a}_2 \rangle := \vt{a}_1 C \vt{a}_2^T \neq 0,
\nn \\
\vt{k}(x,y;t) &=
 \vt{k}_1 (x,t) \e^{\i \la_1 y + \i \la_1^3 t}
   + \vt{k}_2 (x,t) \e^{\i \la_2 y + \i \la_2^3 t}
\nn
\end{align}
into (\ref{Gelf}) and solving it 
with respect to 
$\vt{k}_1$ and $\vt{k}_2$, 
we obtain the ``unrefined'' 
one-soliton solution of 
system (\ref{cdmkdv1}), 
\begin{align}
\vt{u}(x,t) &=
 \vt{k}_1 (x,t) \e^{\i \la_1 x + \i \la_1^3 t}
   + \vt{k}_2 (x,t) \e^{\i \la_2 x+ \i \la_2^3 t}
\nn \\
&=
\frac{\vt{a}_1\e^{\i \la_1 x + \i \la_1^3 t}
+\vt{a}_2\e^{\i \la_2 x+ \i \la_2^3 t}}
{1-\frac{\i (\la_1-\la_2)}{2 (\la_1+\la_2)^2} 
\sca{\vt{a}_1 C}{\vt{a}_2}
\e^{\i (\la_1+\la_2) x + \i (\la_1^3 +\la_2^3)t}}.
\label{onesol1}
\end{align}
Note that the denominator in the above expression 
may become zero for certain values of $x$ and $t$. 
By introducing a new parametrization, 
\[
-\frac{\i (\la_1-\la_2)}{2 (\la_1+\la_2)^2} 
\sca{\vt{a}_1 C}{\vt{a}_2} =: \e^{-2 \delta} \;\,(\delta \in {\mathbb C}), 
\hspace{5mm}
\vt{a}_1 =: 2 \e^{-\delta} \vt{b}_1, \hspace{5mm}
\vt{a}_2 =: 2 \e^{-\delta} \vt{b}_2, 
\]
(\ref{onesol1}) can be rewritten as 
\begin{align}
\vt{u}(x,t) 
&=
\frac{\vt{b}_1\e^{\frac{\i}{2}(\la_1-\la_2) x + \frac{\i}{2} 
(\la_1^3-\la_2^3) t}
+\vt{b}_2 \e^{-\frac{\i}{2}(\la_1-\la_2) x - \frac{\i}{2} (\la_1^3-\la_2^3) t} }
{\cosh  \left[ 
\frac{\i}{2} (\la_1+\la_2) x + \frac{\i}{2} 
(\la_1^3 +\la_2^3)t -\delta \right]
},
\label{onesol2}
\end{align}
with the condition \mbox{$-2\i (\la_1-\la_2) 
\sca{\vt{b}_1 C}{\vt{b}_2} = (\la_1+\la_2)^2$}. 
The ``one-soliton'' solution (\ref{onesol2}) 
resembles 
the 
soliton solution
of the vector mKdV equation (\ref{vmKdV1})~\cite{Tsuchida1}, 
\[
\vt{q}(x,t)= \frac{\vt{c}_1 \e^{\frac{\i}{2} (\la_1-\la_2) x 
+ \frac{\i}{2} (\la_1^3-\la_2^3) t}+
\vt{c}_2 \e^{- \frac{\i}{2} (\la_1-\la_2) x - \frac{\i}{2}(\la_1^3-\la_2^3) t}}
{\cosh \left[ \frac{\i}{2}(\la_1+\la_2) x
  + \frac{\i}{2} (\la_1^3+\la_2^3) t - \delta \right] },
\]
under the conditions 
\mbox{$\sca{\vt{c}_1}{\vt{c}_1}=\sca{\vt{c}_2}{\vt{c}_2}=0$} and 
\mbox{$-8 \sca{\vt{c}_1}{\vt{c}_2} = (\la_1+\la_2)^2$}. 
Moreover, if we impose the ``reality conditions'' 
\mbox{$\la_2 = -\la_1^\ast$} and 
\mbox{$\e^{2\delta} \notin {\mathbb R}_{<0}$}, 
(\ref{onesol2}) provides 
the bright 
one-soliton solution 
of system (\ref{cdmkdv1}) 
that behaves regularly for real $x$ and $t$. 
With the parametrization \mbox{$\la_1=\xi_1 + \i \eta_1$}, 
\mbox{$\la_2=-\la_1^\ast= -\xi_1 + \i \eta_1$} 
(\mbox{$\xi_1 \neq 0$}, \mbox{$\eta_1 > 0$}), 
it reads as 
\begin{align}
\vt{u}(x,t) &=
\frac{\vt{b}_1 \e^{\i \xi_1  x + \i \xi_1 (\x_1^2-3\eta_1^2)  t} +
\vt{b}_2 \e^{-\i \xi_1  x - \i \xi_1 (\x_1^2-3\eta_1^2)  t}}
{\cosh \left[\eta_1 x
  + \eta_1 (3\xi_1^2 - \eta_1^2) t + \delta \right] }, 
\hspace{5mm} 
\i \xi_1 \sca{\vt{b}_1 C}{\vt{b}_2} = \eta_1^2, 
\hspace{5mm} 
{\rm Im}\hspace{1pt}(2\delta) \not\equiv \pi \;({\rm mod}\, 2\pi).
\nn
\end{align}

Similarly, the general $N$-soliton solution 
before imposing the ``reality conditions'' 
is obtained 
by substituting the expressions 
\begin{align}
\vt{f}(x,t) &=
 \sum_{l=1}^{2N} 
\vt{a}_l
\e^{\i \la_l x + \i \la_l^3 t},
\quad
{\rm Im}\, \la_l > 0, 
\quad 
\la_j \neq \la_k \;\, {\rm if}\;\, j \neq k,
\nn \\
\vt{k}(x,y;t) &=
 \sum_{l=1}^{2N} 
 \vt{k}_l (x,t) \e^{\i \la_l 
y + \i \la_l^3 t}
\nn
\end{align}
into (\ref{Gelf}) and solving the resulting linear 
algebraic system for $\vt{k}_l \, (l=1,2, \ldots, 2N)$ 
using a matrix inversion. 
In addition, 
conditions such as 
\[
{\rm Pfaffian \; of \,}\left( 
\frac{\vt{a}_j C \vt{a}_k^T}{\la_j +\la_k}
\right) \neq 0
\]
have to be imposed 
for the solution to decay as \mbox{$x \to - \infty$} 
and to 
behave properly as solitons. The details will be 
published elsewhere. 
It should be noted 
that 
the 
soliton solutions thus obtained 
involve more free parameters, 
and hence, are 
more general than 
the previously known 
solutions~\cite{Iwao2,Loris1}. 

\subsection{Coupled derivative mKdV equations 
of Kaup--Newell type}
\label{KNtype}

\subsubsection{Lax pair for the second flow of 
the matrix Kaup--Newell hierarchy}
Another 
DNLS
equation 
\mbox{$\i \QK_{t_2} + \QK_{xx} \pm \i (|\QK|^2 \QK)_x = 0$}, 
which is often 
referred to as 
the Kaup--Newell equation~\cite{KN}, permits 
an 
integrable matrix generalization~\cite{Konop4,Nij1,Tsuchida3,Dimakis} 
\begin{equation}
\left\{ \hspace{-1pt}
\begin{array}{l}
 \i \QK_{t_2} + \QK_{xx} -\i (\QK \RK \QK)_x = O, \\[1pt]
 \i \RK_{t_2} - \RK_{xx} -\i (\RK \QK \RK)_x = O.
\end{array}
\right.
\label{mKN}
\end{equation}
Here, $\QK$ and $\RK$ 
are 
\mbox{$l_1 \times l_2$} and \mbox{$l_2 \times l_1$}
matrices, respectively.
In this paper, we are more concerned
with the next higher flow in the matrix Kaup--Newell hierarchy 
that commutes with the first nontrivial flow (\ref{mKN}).
It is written as
(up to a scaling of the time variable)~\cite{Nij1}
\begin{equation}
\left\{
 \hspace{1pt}
\begin{split}
& \QK_{t_3} + \QK_{xxx} - \i \frac{3}{2} (\QK_x \RK \QK + \QK \RK \QK_x)_x
 -\frac{3}{2} (\QK \RK \QK \RK \QK)_x = O, \\[2pt]
& \RK_{t_3} + \RK_{xxx} + \i \frac{3}{2} (\RK_x \QK \RK + \RK \QK \RK_x)_x
 -\frac{3}{2} (\RK \QK \RK \QK \RK)_x = O.
\end{split}
\right.
\label{higherKN}
\end{equation}
Note that 
the lower indices of $t$ are omitted in 
the following. 
The Lax pair
for (\ref{higherKN}) is given by
\bseq
\begin{align}
U = \mbox{} &
\i \z^2 \left[
\begin{array}{cc}
 -I_1  &  \\
    &  I_2 \\
\end{array}
\right]
+ 
\z \left[
\begin{array}{cc}
   &  \QK \\
 \RK  &   \\
\end{array}
\right], 
\label{U_form2}
\\[8pt]
V =\mbox{}& 
\i \z^6 
\left[
\begin{array}{cc}
 -4I_1 &  \\
   & 4I_2  \\
\end{array}
\right]
+\z^5
\left[
\begin{array}{cc}
  & 4\QK \\
 4\RK &  \\
\end{array}
\right]
+\i \z^4
\left[
\begin{array}{cc}
-2\QK\RK  &  \\
  & 2\RK\QK \\
\end{array}
\right]
\nn \\
& \mbox{}
+ \z^3
\left[
\begin{array}{cc}
  & 2\i \QK_x + 2 \QK \RK \QK \\
 -2\i \RK_x + 2 \RK \QK \RK &  \\
\end{array}
\right]
\nn \\
& \mbox{}
+\i \z^2
\left[
\begin{array}{cc}
 -\i (\QK_x \RK - \QK \RK_x) - \frac{3}{2} (\QK \RK)^2 &  \\
  & \i (\RK \QK_x - \RK_x \QK) + \frac{3}{2} (\RK \QK)^2 \\
\end{array}
\right]
\nn \\
& \mbox{} + \z 
\left[
\begin{array}{cc}
 & -\QK_{xx} + \i \frac{3}{2} (\QK_x \RK \QK + \QK \RK \QK_x) 
+ \frac{3}{2} (\QK\RK)^2 \QK \\ 
-\RK_{xx} - \i \frac{3}{2} (\RK_x \QK \RK + \RK \QK \RK_x) 
+ \frac{3}{2} (\RK\QK)^2 \RK &
\end{array}
\right].
\label{V_form4}
\end{align}
\label{LaxKN}
\eseq
Substituting the Lax pair (\ref{LaxKN}) 
in
the zero-curvature condition (\ref{Lax_eq}), 
we
obtain the third-order
matrix Kaup--Newell 
system (\ref{higherKN}) without any contradiction or additional
constraint.

\subsubsection{Reduction}
\label{rKN}
Both the first flow (\ref{mKN}) and
the second flow (\ref{higherKN}) 
of the matrix Kaup--Newell hierarchy 
permit the reduction of the Hermitian conjugation~\cite{Makhankov}  
\mbox{$\RK = A_1 \QK^\dagger A_2$}, where $A_1$ and $A_2$ are 
constant Hermitian matrices:\ 
{\mbox{$A_i^\dagger=A_i$}, \hspace{1pt}\mbox{$A_{i,t}=A_{i,x}=O$}. 
Moreover,
unlike the first flow (\ref{mKN}), 
the second flow (\ref{higherKN})
allows an interesting reduction such that $\RK$ is identically
equal to $C \QK^T$, where $C$ is 
an antisymmetric constant matrix:\
{\mbox{$C^T=-C$}, \hspace{1pt}\mbox{$C_t=C_x=O$}. 
For the reduced system to 
assume 
a concise
form without the imaginary unit and fractions,
the following vector reduction
($l_1=1$, $l_2=M$) is considered: 
\begin{equation}
\QK = (u_{1}, \ldots, u_{M}), \hspace{5mm}
\RK = 2\i C \QC^T= 
2\i
\left(
\begin{array}{c}
 \sum_{k=1}^M C_{1 k} u_{k}  \\
  \vdots \\
 \sum_{k=1}^M C_{M k} u_{k}  \\
\end{array}
\right),
\label{set2}
\end{equation}
which also implies
the simple relation \mbox{$\QK \RK =0$}.
System 
(\ref{higherKN}) is 
then reduced to another 
system of coupled derivative mKdV equations:\ 
\begin{equation}
\frac{\6 u_{i}}{\6 t} + \frac{\6^3 u_{i}}{\6 x^3} 
+ 3 \frac{\6}{\6 x} 
\Biggl[
	\sum_{1 \le j<k \le M} C_{jk} \Bigl( \frac{\6 u_{j}}{\6 x} u_{k} 
	- u_{j} \frac{\6 u_{k}}{\6 x} \Bigr) u_{i} \Biggr]
= 0, \hspace{5mm} i=1, 2,\ldots, M.
\label{cdmkdv2}
\end{equation}
To the best of the author's knowledge, this system 
was reported 
for the first time in~\cite{Gakkai}. 
In contrast to (\ref{cdmkdv1}), 
the partial differentiation 
with respect to $x$ 
acts on the entire 
nonlinear term 
in (\ref{cdmkdv2}). 
The conserved densities of the first 
three ranks 
are given by 
\begin{align}
& \hspace{6.5mm} u_i \;\, (i=1, 2, \ldots, M),
\nn \\[2mm]
& \sum_{1 \le j < k \le M} C_{jk} (u_{j,x} u_{k}-u_{j} u_{k,x}), 
\nn \\
& \sum_{1 \le j < k \le M} C_{jk} 
(u_{j,xx} u_{k,x} 
	- u_{j,x} u_{k,xx})
-2\left[ \sum_{1 \le j < k \le M} C_{jk} (u_{j,x} u_k 
	- u_j u_{k,x}) 
\right]^2.
\nn
\end{align}
Using the vector notation, (\ref{cdmkdv2}) can also be written 
as 
\[
\vt{u}_t + \vt{u}_{xxx} 
+ 3\bigl( \sca{\vt{u}_x C}{\vt{u}} \vt{u} \bigr)_x 
= \vt{0}, \hspace{5mm} C^T = -C. 
\]
The Lax pair for this reduced system 
is 
obtained 
by substituting (\ref{set2}) 
into $\QK$ and $\RK$ 
in (\ref{LaxKN}); 
using
a simple gauge transformation,
it can be
rewritten in the form
%
%
\begin{subequations}
\begin{align}
U = \mbox{} &
\left[
\begin{array}{cc}
 -\lambda  &  \vt{u} \\
 \lambda C\vt{u}^T &  O\\
\end{array}
\right], 
\label{spec2}
\\[8pt]
V =\mbox{}& 
\left[
\begin{array}{c|c}
 \lambda^3 + 2\lambda 
	\sca{\vt{u}_x C}{\vt{u}}_{\vphantom \int_{\vphantom 0}}
	 & -\lambda^2 \vt{u} +\lambda \vt{u}_x 
	- \vt{u}_{xx} 
 -3 \sca{\vt{u}_x C}{\vt{u}}_{\vphantom \int_{\vphantom 0}} \vt{u} \\
\hline
\begin{array}{l}
 -\lambda^3 C\vt{u}^T -\lambda^2 C\vt{u}^T_x 
 \vphantom{C^{\vphantom{\int}^{\vphantom{\int}}}}
\\
 -\lambda C\vt{u}_{xx}^T -3\lambda \sca{\vt{u}_x C}{\vt{u}} C\vt{u}^T 
\end{array}
 &  \lambda^2 C\vt{u}^T \vt{u} 
	+\lambda C ( \vt{u}^T_x  \vt{u}-\vt{u}^T \vt{u}_x) \\
\end{array}
\right],
\end{align}
\label{LaxKN2}
\end{subequations}
where \mbox{$\vt{u}$} is a row vector and 
\mbox{$\lambda:=2\i \z^2 $}. 
In the simplest nontrivial case of 
$M=2$, setting 
\[
C_{12}= \i, \hspace{5mm} u_{1} = \psi, \hspace{5mm} u_{2} = \psi^\ast,
\]
the following single equation is obtained:
\begin{equation}
\psi_t + \psi_{xxx} 
+ 3 \i \bigl[ (\psi_x \psi^\ast - \psi \psi^\ast_x) \psi \bigr]_x 
= 0.
\label{single2}
\end{equation}
Using a 
simple point transformation (cf.\ refs.~\citen{IKWS,Kawata2}), 
(\ref{single2}) can be converted into 
an NLS-type equation 
perturbed by higher order terms. 

\subsubsection{Solution formulas}

A set of formulas for 
the solutions of the matrix Kaup--Newell hierarchy, 
decaying as \mbox{$x \to + \infty$}, 
can be derived by exploiting 
its relationship~\cite{Nij1,Tsuchida3} 
with the matrix Chen--Lee--Liu hierarchy studied in subsection~\ref{CLLtype}. 
It 
is written in 
the form of 
the product 
of two matrices as follows~\cite{Tsuchi05}
(cf.\ ref.~\citen{Kawa} 
for the case of scalar variables):
\begin{subequations}
\begin{align}
q
&=
K(x, x
) \left[ I+L (x,x
) \right],
\\[2pt]
r
&=
\left[ I+ L (x,x
)\right]^{-1} \bar{K}(x, x
),
\\[2pt]
K(x, y) 
&=
 \bar{F} (y)
 + \frac{\i}{2} \int^\infty_x \d s_1 \int^\infty_x \d s_2\hspace{1pt} 
K (x, s_1)
F (s_1+s_2-x) \frac{\6 \bar{F} (s_2+y-x)}{\6 s_{2}},
\hspace{5mm} y \ge x, 
\\
\bar{K}(x, y) 
&=
  F (y)
 -\frac{\i}{2}  \int^\infty_x \d s_1 \int^\infty_x \d s_2\hspace{1pt} 
\frac{\6 \bar{K} (x, s_1)}{\6 s_{1}} 
\bar{F}(s_1+s_2-x) F (s_2+y-x), 
\hspace{5mm} y \ge x, 
\\
L (x, y) 
&=
- \frac{\i}{2}
\int^\infty_x \d s \hspace{1pt} F(s) \bar{F} (s+y-x)
\nn \\
& {\hphantom =}\hspace{4pt}
\mbox{}
 + \frac{\i}{2}\int^\infty_x \d s_1 \int^\infty_x \d s_2\hspace{1pt} 
\frac{\6 L (x, s_1)}{\6 s_{1
}}
F (s_1+s_2-x)
\bar{F} (s_2+y-x),
\hspace{5mm} y \ge x.
\end{align}
\label{Gel3}
\end{subequations}
The time dependence of the 
functions 
as well as the index of 
the unit matrix 
$I$ to 
indicate 
its size 
is suppressed 
in the above formulas. 
Here, $\bar{F}(x)$ and $F(x)$ 
satisfy the corresponding {\it linear} \/uncoupled system of 
matrix PDEs, 
{\it e.g.}, (\ref{lin1}) 
for the 
matrix Kaup--Newell system (\ref{mKN}) 
and (\ref{lin2}) 
for the third-order 
flow (\ref{higherKN}), 
and decay 
rapidly as \mbox{$x \to + \infty$}. 
This set of formulas 
is indeed useful 
and sufficient 
for 
constructing explicit solutions of the matrix Kaup--Newell 
hierarchy. However, its 
shortcoming is 
that 
one of the most 
important properties of the hierarchy 
has not been 
incorporated, 
that is, 
the hierarchy 
allows the 
introduction of the potential variables 
\mbox{$q=: \hat{q}_x$} and \mbox{$r=: \hat{r}_x$}, 
and can be reformulated 
in terms of 
$\hat{q}$ and $\hat{r}$. 
Consequently, the 
elementary function 
solutions of the matrix Kaup--Newell hierarchy 
can be expressed as the partial $x$-derivatives of elementary functions; 
this fact has {\it passed 
unnoticed} \/in the existing 
literature (see refs.~\citen{Nij1,NCQL} for some indirect results). 
To bridge 
this gap, we propose a new set of solution 
formulas that 
accurately 
reflects the feasibility of potentiation 
for the matrix Kaup--Newell hierarchy; 
it assumes 
the following 
form 
($x$-derivatives of single quantities~\cite{Nij1,NCQL}): 
\begin{subequations}
\begin{align}
q
&=
\frac{\6 {\cal K}
(x,x)}{\6 x
},
\label{q_div}
\\[2pt]
r
&=
\frac{\6 \bar{{\cal K}} (x,x)
}{\6 x
},
\label{r_div}
\\[2pt]
{\cal K}(x, y) 
&= 
  -\int_y^\infty \d s_1 \hspace{1pt} \bar{F} (s_1)
 +\frac{\i}{2} \int^\infty_x \d s_1 \int^\infty_x \d s_2 
\hspace{1pt} 
\frac{\6 {\cal K} (x, s_1)}{\6 s_{1
}} F (s_1+s_2-x) \bar{F} (s_2+y-x)
\nn 
\\
&=  \bar{G} (y)
 + \frac{\i}{2} \int^\infty_x \d s_1 \int^\infty_x \d s_2\hspace{1pt} 
\frac{\6 {\cal K} (x, s_1)}{\6 s_{1}}
\frac{\6 G (s_1+s_2-x)}{\6 s_2} 
\frac{\6 \bar{G} (s_2+y-x)}{\6 y},
\hspace{5mm} y \ge x, 
\\
\bar{\cal K}(x, y) 
&= 
  -\int_y^\infty \d s_1 \hspace{1pt} F (s_1)
 -\frac{\i}{2}  \int^\infty_x \d s_1 \int^\infty_x \d s_2 
\hspace{1pt} \frac{\6 \bar{\cal K} (x, s_1)}{\6 s_{1}}
\bar{F}(s_1+s_2-x) F (s_2+y-x)
\nn \\
&= 
  G (y)
 -\frac{\i}{2}  \int^\infty_x \d s_1 \int^\infty_x \d s_2\hspace{1pt} 
\frac{\6 \bar{\cal K} (x, s_1)}{\6 s_{1}}
\frac{\6 \bar{G}(s_1+s_2-x)}{\6 s_{2}} 
\frac{\6 G (s_2+y-x)}{\6 y}, 
\hspace{5mm} y \ge x. 
\end{align}
\label{GLM_KN2}
\end{subequations}
Here, 
the matrices 
$\bar{G}$ and $G$ are the 
primitive functions of 
$\bar{F}$ and $F$, 
respectively, 
that also decay 
as \mbox{$x \to +\infty$}, 
that is, \mbox{$\bar{G}(x) := - \int_x^{\infty} 
\bar{F}(y) \hspace{1pt}\d y $} 
and \mbox{$G(x) := - \int_x^{\infty} F(y) \hspace{1pt}\d y $}. 
Note that $\partial/ \partial x$ in (\ref{q_div}) and (\ref{r_div}) 
denotes the partial differentiation with respect to $x$, while all the 
time variables are fixed. 
The two sets of formulas (\ref{Gel3}) and (\ref{GLM_KN2}) 
are indeed equivalent, though it is a highly nontrivial task to 
verify it directly. 

Similarly to the case of 
the matrix Chen--Lee--Liu hierarchy, 
the reduction 
(\ref{set2}) can be realized 
by
setting
\begin{align}
\bar{G}(x,t) = (g_1, g_2, \ldots, g_M)(x,t) =: \vt{g}(x,t),
\hspace{5mm}
G(x,t) = 2\i C \bar{G}(x,t)^T.
\label{GbaG}
\end{align}
Thus, 
the set of
formulas (\ref{GLM_KN2}) is 
reduced to a compact form.
In particular, 
the solutions
to the coupled derivative mKdV equations (\ref{cdmkdv2}), 
decaying as \mbox{$x \to + \infty$}, 
can be constructed
from
those of the
linear vector PDE
\mbox{$\vt{g}_t + \vt{g}_{xxx}=\vt{0}$}
through the formula
\begin{subequations}
\begin{align}
\vt{u}(x,t)
&=
\frac{\6}{\6 x
}
\vt{k}(x, x
;t
),
\\[1pt]
\vt{k}(x, y)
&=
\vt{g}(y)
- 
\int^\infty_x \d s_1 \int^\infty_x \d s_2\hspace{1pt}
\frac{\6 \vt{k}(x, s_1)}{\6 s_1} C 
\frac{\6 \vt{g} (s_1+s_2-x)^T}{\6 s_2}
\frac{\6 \vt{g} (s_2+y-x)}{\6 y}, 
\hspace{5mm} y \ge x.
\label{Gel4}
\end{align}
\label{Gelf2}
\end{subequations}
Here, \mbox{$\vt{u}
= (u_1, u_2, \ldots, u_M)$} and 
$\vt{k}
(x,y)$ are $M$-component row 
vectors. 
Substituting the expressions 
\begin{align}
\vt{g} (x,t) &= \vt{a}_1
\e^{\i \la_1 x + \i \la_1^3 t} + \vt{a}_2 
\e^{\i \la_2 x + \i \la_2^3 t}, 
\quad
{\rm Im}\, \la_j>0 \;(j=1,2), 
\quad
\la_1 \neq \la_2, 
\quad 
\langle \vt{a}_1 C, \vt{a}_2 \rangle := \vt{a}_1 C \vt{a}_2^T \neq 0,
\nn \\
\vt{k}(x,y;t) &=
 \vt{k}_1 (x,t) \e^{\i \la_1 y + \i \la_1^3 t}
   + \vt{k}_2 (x,t) \e^{\i \la_2 y + \i \la_2^3 t}
\nn
\end{align}
into (\ref{Gelf2}) and solving it 
with respect to 
$\vt{k}_1$ and $\vt{k}_2$, 
we obtain the 
``unrefined'' 
one-soliton solution of 
system (\ref{cdmkdv2}), 
\begin{align}
\vt{u}(x,t) &=
\frac{\6}{\6 x}
\left[
 \vt{k}_1 (x,t) \e^{\i \la_1 x + \i \la_1^3 t}
   + \vt{k}_2 (x,t) \e^{\i \la_2 x+ \i \la_2^3 t}
\right]
\nn \\
&=
\frac{\6}{\6 x}
\left[
\frac{\vt{a}_1\e^{\i \la_1 x + \i \la_1^3 t}+\vt{a}_2\e^{\i \la_2 x+ \i \la_2^3 t}}
{1+\frac{\i \la_1 \la_2 (\la_1-\la_2)}{2 (\la_1+\la_2)^2}
\sca{\vt{a}_1 C}{\vt{a}_2}
\e^{\i (\la_1+\la_2) x + \i (\la_1^3 +\la_2^3)t}}
\right].
\label{onesol3}
\end{align}
Note that 
the above solution 
also 
decays 
as \mbox{$x \to - \infty$}, but 
may 
have singularities for certain 
values of $x$ and $t$.
By introducing a new parametrization,
\[
\frac{\i \la_1 \la_2 (\la_1-\la_2)}{2 (\la_1+\la_2)^2}
\sca{\vt{a}_1 C}{\vt{a}_2} =: \e^{-2 \delta} \;\,(\delta \in {\mathbb C}),
\hspace{5mm}
\vt{a}_1 =: 2 \e^{-\delta} \vt{b}_1, \hspace{5mm}
\vt{a}_2 =: 2 \e^{-\delta} \vt{b}_2,
\]
(\ref{onesol3}) can be rewritten as
\begin{equation}
\vt{u}(x,t)
=
\frac{\6}{\6 x}
\left\{
\frac{\vt{b}_1\e^{\frac{\i}{2}(\la_1-\la_2) x + \frac{\i}{2} (\la_1^3-\la_2^3) t}
+\vt{b}_2 \e^{-\frac{\i}{2}(\la_1-\la_2) x - \frac{\i}{2} (\la_1^3-\la_2^3) t} }
{\cosh  \left[
\frac{\i}{2} (\la_1+\la_2) x + \frac{\i}{2}
(\la_1^3 +\la_2^3)t -\delta \right]} \right\},
\label{onesol4}
\end{equation}
with the condition \mbox{$2\i \la_1 \la_2 (\la_1-\la_2)
\sca{\vt{b}_1 C}{\vt{b}_2} = (\la_1+\la_2)^2
$}. 
Moreover, if we impose the ``reality conditions'' 
\mbox{$\la_2 = -\la_1^\ast$} and
\mbox{$\e^{2\delta} \notin {\mathbb R}_{<0}$},
(\ref{onesol4}) provides 
the bright one-soliton solution of system (\ref{cdmkdv2})
that behaves regularly for real $x$ and $t$.
With the parametrization \mbox{$\la_1=\xi_1 + \i \eta_1$},
\mbox{$\la_2=-\la_1^\ast= -\xi_1 + \i \eta_1$}
(\mbox{$\xi_1 \neq 0$}, \mbox{$\eta_1 > 0$}),
it reads as
\begin{align}
\vt{u}(x,t) &=
\frac{\6}{\6 x}
\left\{
\frac{\vt{b}_1 \e^{\i \xi_1  x + \i \xi_1 (\x_1^2-3\eta_1^2)  t} +
\vt{b}_2 \e^{-\i \xi_1  x - \i \xi_1 (\x_1^2-3\eta_1^2)  t}}
{\cosh \left[\eta_1 x
  + \eta_1 (3\xi_1^2 - \eta_1^2) t + \delta \right] } 
\right\}
\nn \\
&= 
\frac{\i \sqrt{\xi_1^2 + \eta_1^2}}
{\cosh^2 \left[\eta_1 x
  + \eta_1 (3\xi_1^2 - \eta_1^2) t + \delta \right]} 
\left\{ 
\vt{b}_1 
\cosh \left[\eta_1 x 
+ \eta_1 (3\xi_1^2 - \eta_1^2) t + \delta +\i \varphi \right]
\e^{\i \xi_1  x + \i \xi_1 (\x_1^2-3\eta_1^2)  t} 
\right.
\nn \\
&
\hspace{45mm}
\left.
\mbox{} -
\vt{b}_2 
\cosh \left[\eta_1 x 
+ \eta_1 (3\xi_1^2 - \eta_1^2) t + \delta -\i \varphi \right]
\e^{-\i \xi_1  x - \i \xi_1 (\x_1^2-3\eta_1^2)  t} \right\},
\nn
\end{align}
with the conditions 
\mbox{$\i \xi_1 \sca{\vt{b}_1 C}{\vt{b}_2} = 
\eta_1^2/(\xi_1^2+ \eta_1^2)$}, 
\mbox{$\exp (\i \varphi)
:= (\xi_1 + \i \eta_1)/
\sqrt{\xi_1^2 + \eta_1^2}$}, 
and 
\\
\mbox{
${\rm Im}\hspace{1pt}(2\delta) 
\not\equiv \pi \;({\rm mod}\, 2\pi)$}.

\subsection{Massive Thirring-like model with symplectic invariance}

\subsubsection{Derivation}
\label{spMTM}

Using our approach, 
it is possible to 
obtain not only evolutionary systems but also 
non-evolutionary systems 
with 
symplectic invariance. 
To demonstrate 
this, 
let us 
consider the 
application of 
the same type of 
reduction as described 
in subsections~\ref{CLLtype} and \ref{KNtype} 
to the first negative flows of 
the matrix DNLS hierarchies. 
The first negative flow of the matrix Chen--Lee--Liu hierarchy 
reads as~\cite{Nij1,Tsuchida3}
\begin{equation}
\left\{ 
\begin{array}{l}
\i \QC_\tau + m \PHC - \hf \PHC \CHC \QC = O, \\[2pt]
\i \RC_\tau - m \CHC + \hf \RC \PHC \CHC = O, \\[2pt]
\i \PHC_x + m \QC - \hf \PHC \RC \QC =O,   \\[2pt]
\i \CHC_x - m \RC + \hf \RC \QC \CHC =O,
\end{array}
\right.
\label{CMTM}
\end{equation}
while 
the first negative flow of the matrix Kaup--Newell hierarchy 
reads as~\cite{Nij1,Tsuchida3}
\begin{equation}
\left\{ 
\begin{array}{l}
\i \QK_\tau + m \PHK - \hf (\QK \CHK \PHK + \PHK \CHK \QK)= O, \\[2pt]
\i \RK_\tau - m \CHK + \hf (\RK \PHK \CHK + \CHK \PHK \RK)= O, \\[2pt]
\i \PHK_x + m \QK =O,  \\[2pt]
\i \CHK_x - m \RK =O.
\end{array}
\right.
\label{KMTM}
\end{equation}
These are the 
matrix generalizations of massive Thirring-type 
models~\cite{Kuz,KN2,KaMoIno,GIK,NCQL}, where $m$ denotes 
an arbitrary nonzero 
constant 
responsible for the mass terms. 
The massless limit \mbox{$m \to 0$} 
is not considered in this paper; 
hereafter, 
the value of $m$ is set as $1$, 
without loss of generality. 
This is easily achieved 
by rescaling 
$\PHC$, $\CHC$, and $\partial_\tau$. 
The Lax pairs for systems (\ref{CMTM}) and (\ref{KMTM}) 
are already known~\cite{Nij1,Tsuchida3}, and 
thus, 
they have been omitted here. 
Both the systems 
permit the reduction 
\mbox{$\RC = C \QC^T$}, 
\mbox{$\CHC = -C \PHC^T$}, \mbox{$C^T = -C$}, 
which is 
the natural extension of the reduction \mbox{$\RC = C \QC^T$}, 
\mbox{$C^T = -C$} 
for the positive 
flows. 

Let us consider a further reduction to the case of 
row/column 
vector variables. 
For the reduced systems to assume 
a concise 
form without the imaginary unit and fractions, 
we consider the following vector reduction: 
\begin{align}
\QK &= (u_{1}, \ldots, u_{M}), \hspace{5mm} 
\RK = 2\i C \QC^T
= 2\i 
\left(
\begin{array}{c}
 \sum_{k=1}^M C_{1 k} u_{k}  \\
  \vdots \\
 \sum_{k=1}^M C_{M k} u_{k}  \\
\end{array}
\right),
\nonumber
\\[4pt]
\PHK &= 
\i (v_{1}, \ldots, v_{M}), \hspace{5mm} 
\CHK = -2\i C \PHK^T
= 2
\left(
\begin{array}{c}
 \sum_{k=1}^M C_{1 k} v_{k}  \\
  \vdots \\
 \sum_{k=1}^M C_{M k} v_{k}  \\
\end{array}
\right).
\nonumber
\end{align}
Then, because 
\mbox{$
\phi \chi =0$}, 
system (\ref{CMTM}) reduces to 
\begin{equation}
\frac{\6^2 u_{i}}{\6 x \6 \tau} + 
u_{i} -
\Biggl[ \sum_{1 \le j < k \le M} C_{jk} 
\Bigl( \frac{\6 u_{j}}{\6 \tau} u_{k} - u_{j} \frac{\6 u_{k}}{\6 \tau} \Bigr) 
 \Biggr] u_{i} = 0,
\hspace{5mm} 
i=1, 2, \ldots, M;
\label{cdMTM1}
\end{equation}
system (\ref{KMTM}) collapses to 
\begin{equation}
\frac{\6^2 v_{i}}{\6 \tau \6 x} + 
v_{i} -
\Biggl[ \sum_{1 \le j < k \le M} C_{jk} 
\Bigl( \frac{\6 v_{j}}{\6 x} v_{k} - v_{j} \frac{\6 v_{k}}{\6 x} \Bigr) 
 \Biggr] v_{i} = 0,
\hspace{5mm} 
i=1, 2, \ldots, M, 
\label{cdMTM}
\end{equation}
where 
\mbox{$u_{i}={\6 v_{i}}/{\6 x}$}, 
\mbox{$\, i=1, 2, \ldots, M$}.
These are considered 
non-evolutionary symmetries of 
system (\ref{cdmkdv1}) and system (\ref{cdmkdv2}), respectively. 
It 
can be easily observed 
that 
(\ref{cdMTM1}) and 
(\ref{cdMTM}) are equivalent up to a simple interchange 
of the 
variables. 
Nonetheless, the 
commutativity of the negative and positive 
flows in each hierarchy itself 
is of significance. 
The Lax pair for (\ref{cdMTM}) is given by 
(cf.~(\ref{LaxKN2}))
\begin{equation}
U = \mbox{} 
\left[
\begin{array}{cc}
 -\lambda  &  \vt{v}_x \\
 \lambda C\vt{v}^T_x &  O\\
\end{array}
\right], \hspace{5mm}
V =\mbox{}
\left[
\begin{array}{cc}
 \frac{1}{\lambda_{\vphantom M}} 
& -
\frac{1}{\lambda_{\vphantom M}} 
\vt{v}
\\
C\vt{v}^T 
 &  - C\vt{v}^T \vt{v} 
\end{array}
\right],
\label{LaxMTM}
\end{equation}
%
where \mbox{$\vt{v}=(v_{1}, v_{2}, \ldots, v_{M})$}. 
In addition, it may 
be noted 
that (\ref{cdMTM}) can be rewritten as a system for 
\mbox{$u_i \hspace{1pt}(=v_{i,x})$}. 
Indeed, (\ref{cdMTM}) implies the relation 
\[
\sum_{1 \le i , l \le M} C_{il} 
\frac{\6 u_{i}}{\6 \tau } u_l + 
\Biggl[ 1 - \sum_{1 \le j < k \le M} C_{jk} 
\Bigl( \frac{\6 v_{j}}{\6 x} v_{k} - v_{j} \frac{\6 v_{k}}{\6 x} \Bigr) 
 \Biggr] 
\sum_{1 \le i,l \le M} C_{il} v_{i} u_{l} 
= 0,
\]
from which we obtain 
\begin{equation}
\sum_{1 \le j < k \le M} C_{jk} 
\Bigl(\frac{\6 v_{j}}{\6 x} v_{k} - v_{j} \frac{\6 v_{k}}{\6 x} \Bigr) 
= \frac{ 1- \sqrt{1-4 \sum_{1 \le i , l \le M} C_{il} 
\frac{\6 u_{i}}{\6 \tau } u_l}}{2}, 
\label{uvrel}
\end{equation}
under appropriate 
boundary conditions. 
Substituting (\ref{uvrel}) in 
(\ref{cdMTM}) with a 
simple operation, 
we 
arrive at a closed PDE system 
for $u_i$, 
\begin{equation}
\frac{\6}{\6 x} \left[
\frac{\Ds 
2\frac{\6 u_{i}}{\6 \tau}}
{1+ \sqrt{1 - 4 \sum_{1 \le j < k \le M} C_{jk}
\Bigl( \frac{\6 u_{j}}{\6 \tau} u_{k} - u_{j} \frac{\6 u_{k}}{\6 \tau} 
\Bigr)}} \right]
+ u_{i} = 0,
\hspace{5mm} 
i=1, 2, \ldots, M.
\label{cdMTM3}
\end{equation}
\subsubsection{Solution formulas}
A set of formulas for the solutions of the 
matrix massive Thirring model (\ref{CMTM}) 
with \mbox{$m=1$}, 
decaying 
as \mbox{$x \to + \infty$}, 
is given 
by 
(\ref{Gel1}) 
supplemented with the following: 
\begin{subequations}
\begin{align}
-\i \hspace{1pt}\PHC
 &= J (x,x
),
\\[2pt]
\i \hspace{1pt}\CHC
 &= 
\bar{J} (x,x
),
\\[2pt]
J(x, y)
&=
 -\int_y^\infty \d s \hspace{1pt} \bar{F} (s)
 - \frac{\i}{2}
\int^\infty_x \d s_1 \int^\infty_x \d s_2\hspace{1pt}
J (x, s_1)
\frac{\6 F (s_1+s_2-x)}{\6 s_2} \bar{F} (s_2+y-x),
\nn \\
&
\hspace{110mm} y \ge x,
\label{}
\\[2pt]
\bar{J}(x, y)
&=
 -\int_y^\infty \d s \hspace{1pt} F (s)
 - \frac{\i}{2} \int^\infty_x \d s_1 \int^\infty_x \d s_2
\hspace{1pt}
\frac{\6 \bar{J} (x, s_1)}{\6 s_1}
\frac{\6 \bar{F}(s_1+s_2-x)}{\6 s_2} 
\int^\infty_{s_2+y-x} \d s_3 \hspace{1pt}F (s_3),
\nn \\
&
\hspace{110mm} 
y \ge x.
\end{align}
\label{supplCLL}
\end{subequations}
Similarly, 
a set of formulas for the 
solutions of the matrix 
Kaup--Newell-type Thirring model (\ref{KMTM}) 
with \mbox{$m=1$}, 
decaying as \mbox{$x \to + \infty$}, 
is given by (\ref{GLM_KN2}) together 
with the following two relations: 
\begin{subequations}
\begin{align}
-\i\hspace{1pt} \PHK
 &= {\cal K}(x,x
),
\\
\i \hspace{1pt}\CHK
&= \bar{\cal K}(x,x
).
\end{align}
\label{supplKN}
\end{subequations}
In both cases, 
the {\it linear} \/matrix PDEs satisfied by 
$\bar{F}$ and $F$ are given by 
\begin{align}
\frac{\6^2 \bar{F}
}{\6 \tau \6 x } + \bar{F}
=O, 
\hspace{5mm}
\frac{\6^2 F
}{\6 \tau \6 x} + F
=O,
\nn
\end{align}
and the same 
relation applies 
for 
their primitive functions $\bar{G}$ and $G$. 
Applying the same reduction 
as that 
in the 
positive flow case 
(cf.\ (\ref{GbaG})) 
to the formulas (\ref{GLM_KN2}) and (\ref{supplKN}), 
we obtain 
the solution 
formula for system (\ref{cdMTM}), 
\begin{subequations}
\begin{align}
\vt{v}(x,\tau)
&=
\vt{k}(x, x;\tau),
\\[2pt]
\vt{k}(x, y)
&=
 \vt{g} (y)
-
\int^\infty_x \d s_1 \int^\infty_x \d s_2\hspace{1pt}
\frac{\6 \vt{k}(x, s_1)}{\6 s_1} 
C \frac{\partial \vt{g} (s_1+s_2-x)^T}{\partial s_2}
\frac{\6 \vt{g} (s_2+y-x)}{\6 y}, 
\hspace{5mm} y \ge x.
\end{align}
\label{hyperSp}
\end{subequations}
Here, \mbox{$\vt{v}= (v_1, v_2, \ldots, v_M)$}, and
$\vt{g}$ 
solves the linear vector PDE 
\mbox{$\vt{g}_{x \tau } + \vt{g}=\vt{0}$}. 
Using the formula (\ref{hyperSp}), we can construct the 
soliton solutions 
of system (\ref{cdMTM}) in 
a manner similar to that 
in the positive flow case. 
In particular, the 
one-soliton solution of (\ref{cdMTM}) is given by 
\begin{align}
\vt{v}(x,\tau) = \frac{\vt{b}_1 \e^{\i \xi_1  
x + \i \frac{\x_1}{\xi_1^2 +\eta_1^2} \tau} 
+
\vt{b}_2 \e^{-\i \xi_1  x - \i \frac{\x_1}{\xi_1^2 + \eta_1^2} \tau}}
{\cosh \left( \eta_1 x
  - \frac{\eta_1}{\xi_1^2 + \eta_1^2} \tau + \delta \right) },
\hspace{5mm}
\i \xi_1 \sca{\vt{b}_1 C}{\vt{b}_2} = \frac{\eta_1^2}{\xi_1^2 + \eta_1^2}.
\nn
\end{align}
The additional conditions 
\mbox{$\xi_1 \in {\mathbb R}\hspace{-1pt}-\hspace{-1pt}\{0\}$}, 
\mbox{$\eta_1 > 0$}, 
and \mbox{$\e^{2\delta} \notin {\mathbb R}_{<0}$}
guarantee that this solution is regular for real $x$ and $\tau$. 

\section{Integrable discretizations}
\label{IntDisc}
\setcounter{equation}{0}

In this section, we propose space discretizations of 
systems 
(\ref{cdmkdv1}), (\ref{cdmkdv2}), and (\ref{cdMTM}) 
while 
retaining both the integrability and symplectic invariance 
in the continuous case. 
Using a tricky 
transformation peculiar to the discrete case, 
we also obtain 
a novel integrable semi-discretization of the Manakov model (\ref{cNLS}). 
One-soliton solutions are 
obtained 
by solving discrete integral
(linear summation) 
equations of the Gel'fand--Levitan--Marchenko type. 

\subsection{System of coupled derivative mKdV equations $(\ref{cdmkdv1})$}
\label{latCLL}

A 
space-discrete analogue of the 
(non-reduced) 
matrix complex mKdV equation (\ref{cmkdv}) 
that is ``maximally symmetric'' 
with respect to 
reduction of the matrix 
variables 
is given by~\cite{Tsuchida4}
\begin{equation}
\left\{ 
\begin{split}
Q_{n,t} &+ (I_1 - Q_{n+1} R_{n+1})^{-1} Q_{n+1}
+ (I_1 - Q_{n} R_{n})^{-1} Q_{n}
\\
& \mbox{}
- (I_1 - Q_{n}R_{n+1})^{-1} Q_{n}
- (I_1 - Q_{n-1}R_{n})^{-1} Q_{n-1} = O \vspace{1.5mm},
\\[3pt]
R_{n,t} &+ R_{n+1}(I_1 - Q_{n} R_{n+1})^{-1} 
+ R_{n}(I_1 - Q_{n-1} R_{n})^{-1} 
\\
& \mbox{}
- R_{n} (I_1 - Q_{n}R_{n})^{-1} 
- R_{n-1} (I_1 - Q_{n-1}R_{n-1} )^{-1} = O. 
\end{split}
\right.
\label{sdmKdV}
\end{equation}
Here, 
``maximally symmetric'' implies 
that (\ref{sdmKdV}) 
possibly permits 
all the 
interesting reductions 
corresponding to those 
in the continuous case. 
System (\ref{sdmKdV}) 
possesses the following Lax pair:
%
\begin{subequations}
\begin{align}
L_n &= \left[
\begin{array}{cc}
z I_1 - \left( z+\frac{1}{z} \right) Q_n R_n
        & \frac{1}{z} Q_n \\[2pt]
 -\left( z + \frac{1}{z} \right) R_n  &  \frac{1}{z} I_2 \\
\end{array}
\right]
=
\left[
\begin{array}{cc}
z I_1 & Q_n \\
 O &  I_2 \\
\end{array}
\right]
\left[
\begin{array}{cc}
 I_1 & O \\
 -\left( z + \frac{1}{z} \right) R_n  &  \frac{1}{z} I_2 \\
\end{array}
\right], 
\label{Ln_KN}
\\[8pt]
 M_n &=
\left[
\begin{array}{c|c}
\begin{array}{l}
-\left( z^2 +\frac{1}{z^2} +2 \right) I_1
\\[2pt]
 \mbox{} + \left( 1+\frac{1}{z^2} \right)_{\vphantom M_{\vphantom M}}
\hspace{-1pt}
	(I_1 - Q_{n-1}R_n)^{-1}
\end{array}
&
\begin{array}{l}
 -(I_1 - Q_n R_n)^{-1} Q_n
\\[1pt]
 \mbox{} - \frac{1}{z^2_{\vphantom M}} (I_1 - Q_{n-1} R_n)^{-1} Q_{n-1} 
\end{array}
\\
\hline
\begin{array}{l}
 \left( z^{2}+1 \right)^{\vphantom M} 
(I_2- R_{n-1} Q_{n-1})^{-1} R_{n-1}
\\[1pt]
 \mbox{}+ \left( 1+ \frac{1}{z^2} \right) (I_2 - R_n Q_{n-1})^{-1} R_n 
\end{array}
&
- \left( 1 + \frac{1}{z^2} \right) (I_2 - R_n Q_{n-1})^{-1}
\end{array}
\right]. 
\label{Mn_KN}
\end{align}
\label{Lax_KN}
\end{subequations}
Here, $z$ is the
spectral parameter independent of $n$ and $t$; 
$I_1$ and $I_2$ are
the \mbox{$l_1 \times l_1$} and
\mbox{$l_2 \times l_2$} unit
matrices, respectively.
Indeed, substituting (\ref{Lax_KN}) in
(a 
space-discrete version of) the zero-curvature
condition~\cite{Kako} 
\begin{equation}
 L_{n,t} +L_n M_n - M_{n+1}L_n = O,
\label{sdLax_eq}
\end{equation}
which is the compatibility condition for
the overdetermined linear 
equations 
%
\mbox{$\Psi_{n+1} = L_n \Psi_n$}
and 
\mbox{$\Psi_{n,t} = M_n \Psi_n$},
we obtain the  space-discrete 
system (\ref{sdmKdV}). 
To achieve 
a discrete counterpart 
of the reduction (\ref{red1}), 
we first assume the following relations 
between 
\mbox{$Q_n \, (n \in {\mathbb Z})$}
and 
\mbox{$R_n \, (n \in {\mathbb Z})$}: 
\begin{equation}
R_n = P_n - P_{n-1}, \quad 
Q_m P_n + Q_n P_m =O, \hspace{5mm} \forall \hspace{1pt}m,n \in {\mathbb Z}. 
\label{constr}
\end{equation}
If these are satisfied, then 
system (\ref{sdmKdV}) is reduced to the form
\begin{equation}
\left\{ 
\begin{split}
& Q_{n,t} + (I_1 + Q_{n+1} P_{n})^{-1} (Q_{n+1} - Q_n)
+ (I_1 + Q_{n} P_{n-1})^{-1} (Q_{n}-Q_{n-1}) =O,
\\ 
& \bigl[ 
P_{n,t} + (P_{n+1}-P_n)(I_1 - Q_{n} P_{n+1})^{-1} 
+ (P_{n}-P_{n-1}) (I_1 - Q_{n-1} P_{n})^{-1} \bigr]
- \bigl[ n \to n-1\bigr]
=O.
\end{split}
\right.
\label{QPeq}
\end{equation}
The following choice of $Q_n$ and $P_n$
automatically satisfies (\ref{constr}):
\begin{equation}
Q_n = (u_n^{(1)}, \ldots, u_n^{(M)}), \hspace{5mm} 
P_n = C Q_n^T
= \left(
\begin{array}{c}
 \sum_{k=1}^M C_{1 k} u_n^{(k)}  \\
  \vdots \\
 \sum_{k=1}^M C_{M k} u_n^{(k)}  \\
\end{array}
\right), 
\hspace{5mm} C^T = -C. 
\label{setting3}
\end{equation}
Substituting 
(\ref{setting3}) in 
(\ref{QPeq}), 
we obtain an integrable semi-discretization of the coupled derivative 
mKdV equations (\ref{cdmkdv1}), 
\begin{align}
& \frac{\6 u_n^{(i)}}{\6 t} + 
\frac{u_{n+1}^{(i)} - u_{n}^{(i)}}{\Ds 1+ \sum_{1 \le j<k \le M} C_{jk} 
	\bigl( u_{n+1}^{(j)} u_{n}^{(k)} - u_{n}^{(j)} u_{n+1}^{(k)}\bigr)} +
\frac{u_{n}^{(i)} - u_{n-1}^{(i)}}{\Ds 1+ \sum_{1 \le j<k \le M} C_{jk} 
	\bigl(u_{n}^{(j)} u_{n-1}^{(k)} - u_{n-1}^{(j)} u_{n}^{(k)}\bigr) } =0,
\nn \\
& \hspace{110mm} i=1,2, \ldots, M. 
\label{discrete1}
\end{align}
The Lax pair for 
the 
reduced system (\ref{discrete1}) is obtained by substituting (\ref{setting3}) 
into $Q_n$ and \mbox{$R_n (=P_n - P_{n-1})$} in (\ref{Lax_KN}).
Moreover, using a 
gauge transformation, 
we can restore the ultralocality of the 
spatial Lax matrix $L_n$; 
thus, the Lax pair
can be
written in the form
\begin{subequations}
\begin{align}
L_n &= \left[
\begin{array}{cc}
 \mu & \left( \mu +1 \right) \vt{u}_n 
\\[2pt]
 \left( \mu - 1 \right) C \vt{u}_n^T 
 & I + \left( \mu +1 \right)
	C\vt{u}_n^T \vt{u}_n \\
\end{array}
\right], 
\label{d-spec1}
\\[8pt]
 M_n &=  \frac{-1}{1+\sca{\vt{u}_n C}{\vt{u}_{n-1}}}
\left[
\begin{array}{c|c}
 \mu -\frac{1}{\mu_{\vphantom M}} 
&
 \left( \mu+1 \right) \vt{u}_n 
 + \left( 1+ \frac{1}{\mu}\right)_{\vphantom M} 
\vt{u}_{n-1} 
\\
\hline
\begin{array}{l}
  \left( \mu -1 \right) C\vt{u}_{n-1}^{\hspace{1pt}T^{\vphantom M}}
\\[1pt] 
\mbox{}
 + \left( 1- \frac{1}{\mu} \right) C\vt{u}_{n}^T
\end{array}
&
\begin{array}{l}
 \left( \mu +1 \right) C\vt{u}_{n-1}^{\hspace{1pt}T^{\vphantom M}} \vt{u}_{n}
\\[1pt] 
\mbox{}+ \left( 1 + \frac{1}{\mu} \right) C\vt{u}_{n}^T \vt{u}_{n-1}
\end{array}
\end{array}
\right],
\end{align}
\end{subequations}
where \mbox{$\vt{u}_n = (u_n^{(1)}, u_n^{(2)}, \ldots, u_n^{(M)})$} 
is a row vector 
and \mbox{$\mu :=z^2 $} 
is the ``new'' spectral parameter. 
It is easy to see that 
the discrete eigenvalue 
problem \mbox{$\Psi_{n+1}=L_n \Psi_n$} 
with (\ref{d-spec1}) 
reduces to 
the continuous eigenvalue 
problem 
\mbox{$\Psi_x=U\Psi$} with (\ref{spec1}) in a 
suitable 
continuous limit (cf.\ ref.~\citen{Ab78}). 

\subsection{Manakov model $(\ref{cNLS})$}
Let us consider the canonical case 
of the coupling constants 
(cf.\ the introduction) 
in (\ref{discrete1}), namely, 
\mbox{$C_{2j-1 \hspace{1pt}2k}=-C_{2k \hspace{1pt}2j-1}= 
\de_{j k}$}, 
\mbox{$C_{2j-1 \hspace{1pt}2k-1}=C_{2j \hspace{1pt}2k}=0$}, 
and \mbox{$M=2m$}, 
and 
change the dependent 
variables as follows: 
\begin{align}
u_n^{(2j-1)} =: (-\i)^n \e^{2\i t} q_n^{(j)}, \hspace{5mm}
u_n^{(2j)} =: \i^{n+1} \e^{-2\i t} r_n^{(j)}, \hspace{5mm}
j=1, 2, \ldots, m.
\nonumber
\end{align}
Then, 
system (\ref{discrete1}) is converted 
into the following 
form: 
\[
\left\{ 
\begin{split}
\i \frac{\6 q_n^{(j)}}{\6 t} + 
\frac{q_{n+1}^{(j)} - \i q_{n}^{(j)}}{\Ds 1+ \sum_{k=1}^{m} 
	\bigl( q_{n+1}^{(k)} r_{n}^{(k)} + q_{n}^{(k)} r_{n+1}^{(k)}\bigr)} 
+\frac{q_{n-1}^{(j)} +\i q_{n}^{(j)}}{\Ds 1+ \sum_{k=1}^{m} 
	\bigl(q_{n}^{(k)} r_{n-1}^{(k)} + q_{n-1}^{(k)} r_{n}^{(k)}\bigr) } 
	-2 q_{n}^{(j)}=0, \\[2pt]
\i \frac{\6 r_n^{(j)}}{\6 t} - 
\frac{r_{n+1}^{(j)} + \i r_{n}^{(j)}}{\Ds 1+ \sum_{k=1}^{m} 
	\bigl( q_{n+1}^{(k)} r_{n}^{(k)} + q_{n}^{(k)} r_{n+1}^{(k)}\bigr)} 
 - \frac{r_{n-1}^{(j)} -\i r_{n}^{(j)}}{\Ds 1+ \sum_{k=1}^{m} 
	\bigl(q_{n}^{(k)} r_{n-1}^{(k)} + q_{n-1}^{(k)} r_{n}^{(k)}\bigr) } 
	+2 r_{n}^{(j)}=0, \\
j=1,2, \ldots, m. 
\end{split}
\right. 
\]
By further 
imposing the reduction 
\mbox{$r_{n}^{(j)} = \sigma_j
q_{n}^{(j)\hspace{1pt}\ast}$}, \mbox{$\sigma_j= \pm 1$}, 
we obtain a new integrable 
semi-discretization 
of the 
system of 
coupled 
NLS equations. 
The simplest case 
\mbox{$r_{n}^{(j)} = - q_{n}^{(j)\hspace{1pt}\ast}$} 
\mbox{$(j=1, 2, \ldots, m)$}
provides 
a space-discrete analogue of the $U(m)$-invariant 
Manakov model (\ref{cNLS}), 
\begin{align}
 {\rm i} \vt{q}_{n,t}
 +\frac{\vt{q}_{n+1}-\i \vt{q}_n}
{1-\sca{\vt{q}_{n+1}}{\vt{q}_n^\ast}-\sca{\vt{q}_{n}}{\vt{q}_{n+1}^\ast}} 
 +\frac{\vt{q}_{n-1}+\i \vt{q}_n}
{1-\sca{\vt{q}_{n}}{\vt{q}_{n-1}^\ast}-\sca{\vt{q}_{n-1}}{\vt{q}_{n}^\ast}} 
  - 2 \vt{q}_n = \vt{0},
\label{sdcNLS}
\end{align}
where $\vt{q}_n = \bigl( q_n^{(1)}, q_n^{(2)}, \ldots, q_n^{(m)} \bigr)$. 

\subsection{Solutions 
to system $(\ref{discrete1})$}
\label{}
A set of formulas for the solutions of the 
space-discrete matrix complex mKdV 
system (\ref{sdmKdV}) 
as well as its commuting 
flows, which 
tend to zero 
as \mbox{$n \to + \infty$}, is given by 
\begin{subequations}
\begin{align}
Q_n &= K(n,n)
+ K(n+1,n+1), 
\\[2pt]
R_n &= \mbox{} \bar{K} (n,n)
+ \bar{K} (n+1,n+1),
\\[2pt]
K (n,m)
&= \bar{G}(m)
+ \sum_{j=0}^{\infty} \sum_{k=0}^{\infty}
\left[ K(n,n+j) + K(n,n+j+1)\right]
\left[ G(n+j+k+1) + G(n+j+k+2) \right]
\nn \\
& {\hphantom =}
\hspace{25mm}
\mbox{}\times
\left[ \bar{G}(m+k) + \bar{G}(m+k+1) \right],
\hspace{5mm} m \geq n,
\\[2pt]
\bar{K} (n,m)
&= - G(m) +
\sum_{j=0}^{\infty}
\sum_{k=0}^{\infty}
\left[ \bar{K} (n,n+j) + \bar{K}(n,n+j+1) \right]
\left[ \bar{G}(n+j+k) + \bar{G}(n+j+k+1)
\right]
\nn \\
& {\hphantom =}
\hspace{25mm}
\mbox{} \times
\left[ G(m+k) + G(m+k+1) \right],
\hspace{5mm} 
m \geq n.
\end{align}
\label{mKdVGLM}
\end{subequations}
Here, the functions $\bar{G}(n)$ and $G(n)$ 
satisfy 
the corresponding {\it linear} \/uncoupled 
system of matrix 
differential-difference equations, 
{\it e.g.}, 
\begin{align}
\frac{\6 \bar{G}(n)}{\6 t} + \bar{G}(n+1) - \bar{G}(n-1)=O,
\hspace{5mm}
\frac{\6 G(n)}{\6 t} + G(n+1) - G(n-1)=O
\nn
\end{align}
for the flow 
(\ref{sdmKdV}), and decay rapidly as \mbox{$n \to + \infty$}. 
The reduction 
given by (\ref{constr}) and (\ref{setting3}) is achieved 
at the level of the 
solution formulas 
by 
setting 
\begin{align}
\bar{G}(n) = (g_1, g_2, \ldots, g_M)(n) =: \vt{g}(n),
\hspace{5mm}
G(n) = -C \bigl[\bar{G}(n) - \bar{G}(n-1)\bigr]^T.
\nn
\end{align}
With this reduction, the 
set of solution formulas 
(\ref{mKdVGLM}) is 
reduced to a compact form. 
Thus, the solutions to the 
semi-discrete 
coupled derivative mKdV equations (\ref{discrete1}), 
decaying as \mbox{$n \to + \infty$}, can be constructed 
from those of the 
linear vector differential-difference equation 
\mbox{$\partial\vt{g}(n)/\partial t+\vt{g}(n+1) - \vt{g}(n-1)=\vt{0}$} 
through the formula
\begin{subequations}
\begin{align}
\vt{u}_n(t)
&=
\vt{k}(n, n;t) 
+ \vt{k}(n+1, n+1
;t
),
\\[2pt]
\vt{k}(n, m) 
&= 
 \vt{g} (m) +
\sum_{j=0}^\infty \sum_{l=0}^\infty 
\bigl[ \vt{k}(n, n+j) + \vt{k}(n, n+j+1) \bigr]
C \bigl[ \vt{g} (n+j+l) - \vt{g} (n+j+l+2) 
\bigr]^T
\nn \\ 
& {\hphantom =} \hspace{21mm}\mbox{}
\times 
\bigl[ \vt{g} (m+l) + \vt{g} (m+l+1) \bigr], \hspace{5mm} m \ge n.
\label{}
\end{align}
\label{sdlinear1}
\end{subequations}
Here, \mbox{$\vt{u}_n = (u_n^{(1)}, u_n^{(2)}, \ldots, u_n^{(M)})$} and 
\mbox{$\vt{k}(n,m)$}
are $M$-component row vectors. 
Substituting the expressions 
\begin{align}
\vt{g} (n,t) &= \vt{a}_1
\mu_1^{-n} \e^{(\mu_1-\mu_1^{-1}) t} 
+ \vt{a}_2 \mu_2^{-n} \e^{(\mu_2-\mu_2^{-1}) t}, 
\quad
|\mu_j|>1 \;(j=1,2), 
\quad
\mu_1 \neq \mu_2, 
\quad 
\langle \vt{a}_1 C, \vt{a}_2 \rangle 
\neq 0,
\nn \\
\vt{k}(n,m;t) &=
 \vt{k}_1 (n,t) \mu_1^{-m} \e^{(\mu_1-\mu_1^{-1}) t} 
   + \vt{k}_2 (n,t) \mu_2^{-m} \e^{(\mu_2-\mu_2^{-1}) t}
\nn
\end{align}
into (\ref{sdlinear1}) and solving it 
with respect to 
$\vt{k}_1$ and $\vt{k}_2$, 
we obtain the 
``unrefined'' 
one-soliton solution of 
system (\ref{discrete1}) in the additive 
form
\begin{equation}
\vt{u}_n (t) = \vt{p}_n (t) + \vt{p}_{n+1} (t),
\label{sumform}
\end{equation}
where $\vt{p}_n(t)$ is given by 
\begin{align}
\vt{p}_n (t) &=
 \vt{k}_1 (n,t) \mu_1^{-n} \e^{(\mu_1-\mu_1^{-1}) t} 
   + \vt{k}_2 (n,t) \mu_2^{-n} \e^{(\mu_2-\mu_2^{-1}) t} 
\nn \\
&= 
\frac{\vt{a}_1 \mu_1^{-n} \e^{ (\mu_1-\mu_1^{-1})t}
+\vt{a}_2 \mu_2^{-n} \e^{ (\mu_2 -\mu_2^{-1})t}}
{1+ \frac{(\mu_1-\mu_2)(1+\mu_1)(1+\mu_2)}{(1-\mu_1 \mu_2)^2}
\sca{\vt{a}_1 C}{\vt{a}_2}
\mu_1^{-n} \mu_2^{-n} \e^{ (\mu_1 +\mu_2-\mu_1^{-1}-\mu_2^{-1})t}}.
\label{onesol5}
\end{align}
In fact, the row vector 
$\vt{p}_n (t)$ itself 
satisfies the 
nonlinear 
differential-difference 
equation (cf.\ (\ref{discrete1}) and (\ref{sumform})),
\begin{equation}
\frac{\6 \vt{p}_n}{\6 t} + 
\frac{\vt{p}_{n+1}-\vt{p}_{n-1}}{1+ \sca{\vt{p}_{n+1}C}{\vt{p}_{n}}
+ \sca{\vt{p}_{n}C}{\vt{p}_{n-1}} + \sca{\vt{p}_{n+1}C}{\vt{p}_{n-1}}
} = \vt{0}.
\label{peq}
\end{equation}
Note that the denominator in 
the expression (\ref{onesol5}) 
may become zero for certain values of $n$ and $t$.
By introducing a new parametrization,
\begin{align}
& \frac{(\mu_1-\mu_2)(1+\mu_1)(1+\mu_2)}{(1-\mu_1 \mu_2)^2}
\sca{\vt{a}_1 C}{\vt{a}_2} 
=: \e^{-2 \delta} \;\,(\delta \in {\mathbb C}),
\nn \\ 
&
\vt{a}_1 =: 2 \e^{-\delta} \vt{b}_1, \hspace{5mm}
\vt{a}_2 =: 2 \e^{-\delta} \vt{b}_2,
\hspace{5mm} 
\mu_1 = \e^{\alpha - \i \beta},
\hspace{5mm} 
\mu_2 = \e^{\alpha + \i \beta}, 
\nn
\end{align}
(\ref{onesol5}) can be rewritten as
\begin{align}
\vt{p}_n (t)
=
\frac{\vt{b}_1 \e^{\i \beta n -2\i(\cosh \alpha \hspace{1pt}\sin \beta) t}
+\vt{b}_2 \e^{-\i \beta n + 2\i(\cosh \alpha \hspace{1pt}\sin \beta) t}}
{\cosh  \left[
\alpha n -2 (\sinh \alpha \hspace{1pt}\cos \beta) t +\delta
 \right]
},
\label{onesol6}
\end{align}
with the condition 
\mbox{$-4\i (\cosh \alpha + \cos \beta)\sin \beta \hspace{2pt}
\sca{\vt{b}_1 C}{\vt{b}_2} = (\sinh \alpha)^2 $}.
The ``one-soliton'' solution (\ref{onesol6})
of (\ref{peq})
resembles
the
soliton solution
of the semi-discrete vector mKdV equation 
(or, 
the 
vector modified Volterra lattice)
\mbox{$\partial \vt{q}_{n}/\partial t
= \bigl(1+ \sca{\vt{q}_n}{\vt{q}_n}\bigr)
(\vt{q}_{n+1} -\vt{q}_{n-1})$} given by 
\[
\vt{q}_n (t)= \frac{\vt{c}_1 \e^{\i \beta n 
+ 2\i(\cosh \alpha \hspace{1pt}\sin \beta) t}
+ \vt{c}_2 \e^{-\i \beta n -2 \i(\cosh \alpha\hspace{1pt} \sin \beta) t}}
{\cosh \left[ \alpha n
  + 2 (\sinh \alpha \hspace{1pt}\cos \beta) t + \delta \right] },
\]
under the conditions
\mbox{$\sca{\vt{c}_1}{\vt{c}_1}=\sca{\vt{c}_2}{\vt{c}_2}=0$} and 
\mbox{$\hspace{1pt}2 \sca{\vt{c}_1}{\vt{c}_2} = (\sinh \alpha)^2$}.
Moreover, if 
the ``reality conditions''
\mbox{$\alpha >0$}, 
\mbox{$0< \beta < \pi$} 
(or \mbox{$-\pi< \beta < 0$}), 
and
\mbox{$\e^{2\delta} \notin {\mathbb R}_{<0}$} 
are imposed, (\ref{onesol6}) provides 
the bright
one-soliton solution
that is indeed regular 
for real $n$ and $t$. 
Owing to the discrete nature of the space variable $n$, 
there can exist 
other cases 
wherein the solution 
is regular, 
{\em e.g.}, \mbox{$\alpha >0$}, 
\mbox{$\beta=\pi/2$}, 
and \mbox{$\e^{2\alpha n + 2 \delta} \neq -1
$}, 
$\forall \hspace{1pt}n$. 

\subsection{System of coupled derivative mKdV equations $(\ref{cdmkdv2})$}
\label{type2}

In this subsection 
and the subsequent subsections, 
we 
use 
the forward 
difference operator $\boldsymbol{\Delta}_n$ to indicate 
\[
\boldsymbol{\Delta}_n 
f_{n+j} 
:=
f_{n+j+1}
-f_{n+j}.
\]
It can be recalled 
that an integrable semi-discretization of the 
third-order matrix 
Kaup--Newell system 
(\ref{higherKN}) is given by~\cite{Tsuchida4}
\begin{equation}
\left\{ 
\hspace{1pt}
\begin{split}
& \QK_{n,t} + \boldsymbol{\Delta}_n \bigl[(I_1 - \QK_{n}\RK_{n})^{-1} \QK_{n}
+ (I_1 + \QK_{n-1}\RK_{n})^{-1} \QK_{n-1} \bigr] = O,
\\[1pt]
& \RK_{n,t} + \boldsymbol{\Delta}_n \bigl[ \RK_{n} (I_1 + \QK_{n-1} \RK_{n})^{-1} 
+ \RK_{n-1} (I_1 - \QK_{n-1} \RK_{n-1} )^{-1} \bigr] = O. 
\end{split}
\right.
\label{sdKN}
\end{equation}
%
This system possesses the following Lax pair: 
%
\begin{subequations}
\begin{align}
L_n &= \left[
\begin{array}{cc}
z I_1 - \left( z-\frac{1}{z} \right) \QK_n \RK_n
        & \frac{1}{z} \QK_n \\[2pt]
 \left( -z + \frac{1}{z} \right) \RK_n  
	&  \frac{1}{z} I_2 \\
\end{array}
\right]
=
\left[
\begin{array}{cc}
z I_1 & \QK_n \\
 O &  I_2 \\
\end{array}
\right]
\left[
\begin{array}{cc}
 I_1 & O \\
  \left( -z + \frac{1}{z} \right) \RK_n  &  \frac{1}{z} I_2 \\
\end{array}
\right],
\label{Ln_KN2}
\\[8pt]
M_n &=
\left[
\begin{array}{c|c}
\begin{array}{l}
\left[ -\left( z^2 -1 \right) + \left( 1-\frac{1}{z^2} \right) \right] I_1
\\[1pt]
 \mbox{} -\left( 1-\frac{1}{z^2} \right)_{\vphantom \int}
	(I_1 + \QK_{n-1}\RK_n)^{-1}
\end{array}
&
\begin{array}{l}
- (I_1-\QK_n \RK_n)^{-1} \QK_n
\\[1pt]
 \mbox{} - \frac{1}{z^2_{\vphantom M}} (I_1 + \QK_{n-1} \RK_n)^{-1} \QK_{n-1}
\end{array}
\\
\hline
\begin{array}{l}
\left( z^2 - 1 \right)^{\vphantom M}
 (I_2- \RK_{n-1} \QK_{n-1})^{-1}\RK_{n-1}
\\[1pt]
 \mbox{}+ \left( 1 - \frac{1}{z^2} \right)
(I_2 + \RK_n \QK_{n-1})^{-1} \RK_n 
\end{array}
&
 \left( 1 - \frac{1}{z^2} \right) (I_2 + \RK_n \QK_{n-1})^{-1}
\end{array}
\right].
\label{Mn_KN2}
\end{align}
\label{Lax_KN2}
\end{subequations}
Indeed, the 
substitution of 
(\ref{Lax_KN2}) in 
the 
zero-curvature
condition (\ref{sdLax_eq}) 
gives the  space-discrete system (\ref{sdKN}). 
As 
system (\ref{higherKN}) 
permits 
the reduction 
\mbox{$\RC \propto C \QC^T$}, {\mbox{$C^T=-C$}, 
so 
system (\ref{sdKN}) 
allows the 
corresponding 
reduction 
\mbox{$\RK_n = C \QK_{n-\hf}^{\hspace{1pt}T}$}, 
\mbox{$C^T=-C$}. In particular, 
considering the vector reduction
\[
\QK_n = (u_n^{(1)}, \ldots, u_n^{(M)}), \hspace{5mm}
\RK_n = C \QK_{n-\hf}^{\hspace{1pt}T}, \hspace{5mm} C^T= -C,
\]
we obtain an integrable semi-discretization of 
the 
coupled derivative mKdV equations (\ref{cdmkdv2}),
\begin{align}
 \frac{\6 u_n^{(i)}}{\6 t} + \boldsymbol{\Delta}_n 
\Biggl[ 
& \frac{u_{n}^{(i)}}{\Ds 1 - \sum_{1 \le j<k \le M} C_{jk} 
	\bigl( u_{n}^{(j)} u_{n-\hf}^{(k)} - u_{n-\hf}^{(j)} u_{n}^{(k)} 
	\bigr)} 
\nn \\
& \mbox{}+
\frac{u_{n-1}^{(i)}}{\Ds 1 - \sum_{1 \le j<k \le M} C_{jk} 
	\bigl(u_{n-\hf}^{(j)} u_{n-1}^{(k)} - u_{n-1}^{(j)} u_{n-\hf}^{(k)} 
	\bigr) } \Biggr] =0,
\hspace{5mm}
i=1,2, \ldots, M. 
\label{discrete2}
\end{align}
This space 
difference 
scheme 
depends on 
the five 
points:\
$n$, \mbox{$n \pm \hf$}, and \mbox{$n \pm 1$}. 
In fact, 
we can 
derive a simpler, 
three-point difference scheme 
for 
(\ref{cdmkdv2}) 
from the same matrix 
system (\ref{sdKN}). 
For this purpose, we consider the vector 
reduction
\begin{align}
\QK_n = (u_n^{(1)}, \ldots, u_n^{(M)}), \hspace{5mm}
\RK_n = C (\QK_{n} + \QK_{n-1})^T, \hspace{5mm} C^T= -C. 
\label{QRn1}
\end{align}
Then, considering 
the relation 
\mbox{$\QK_n C \QK_m^T + \QK_m C \QK_n^T =0$}, 
we can observe 
that (\ref{sdKN}) is reduced to an alternative lattice version 
of (\ref{cdmkdv2}),
\begin{align}
\frac{\6 u_n^{(i)}}{\6 t} + \boldsymbol{\Delta}_n \Biggl[
\frac{u_{n}^{(i)} + u_{n-1}^{(i)}}{\Ds 1- \sum_{1 \le j<k \le M} C_{jk} 
	\bigl(u_{n}^{(j)} u_{n-1}^{(k)} - u_{n-1}^{(j)} u_{n}^{(k)}\bigr) } 
  \Biggr] = 0, \hspace{5mm} i=1,2, \ldots, M. 
\label{discrete3}
\end{align}
System (\ref{discrete3}) resembles 
the 
integrable semi-discretization of the vector 
third-order Heisenberg ferromagnet model (\ref{hoHF}) 
(see (2.22) in 
ref.~\citen{Tsuchida5} 
or (5.25) in ref.~\citen{GIV}). 
The Lax pair for
(\ref{discrete3}) is obtained by substituting 
(\ref{QRn1}) into $\QK_n$ and \mbox{$\RK_n$} in (\ref{Lax_KN2}). 
Moreover, using a gauge transformation, we can restore 
the ultralocality of the $L_n$-matrix; 
thus, the Lax pair
can be
written in the form
\begin{subequations}
\begin{align}
L_n &= \left[
\begin{array}{cc}
 \mu & \left( \mu +1 \right) \vt{u}_n
\\[2pt]
 \left( -\mu + 1 \right) C \vt{u}_n^T
 & I - \left( \mu -1 \right)
        C\vt{u}_n^T \vt{u}_n \\
\end{array}
\right],
\label{d-spec2}
\\[8pt]
 M_n &=  \frac{1}{1-\sca{\vt{u}_n C}{\vt{u}_{n-1}}}
\left[
\begin{array}{c|c}
 -\mu +\frac{1}{\mu_{\vphantom M}}
&
 -\left( \mu+1 \right) \vt{u}_n
 - \left( 1+ \frac{1}{\mu}\right)_{\vphantom M}
\vt{u}_{n-1}
\\
\hline
\begin{array}{l}
  \left( \mu -1 \right) C\vt{u}_{n-1}^{\hspace{1pt}T^{\vphantom M}}
\\[1pt]
\mbox{}
 + \left( 1- \frac{1}{\mu} \right) C\vt{u}_{n}^T
\end{array}
&
\begin{array}{l}
 \left( \mu -1 \right) C\vt{u}_{n-1}^{\hspace{1pt}T^{\vphantom M}} \vt{u}_{n}
\\[1pt]
\mbox{}- \left( 1 - \frac{1}{\mu} \right) C\vt{u}_{n}^T \vt{u}_{n-1}
\end{array}
\end{array}
\right],
\end{align}
\label{Lax_KN3}
\end{subequations}
where \mbox{$\vt{u}_n = (u_n^{(1)}, u_n^{(2)}, \ldots, u_n^{(M)})$}
and \mbox{$\mu :=z^2 $}. 
The discrete eigenvalue
problem \mbox{$\Psi_{n+1}=L_n \Psi_n$}
with (\ref{d-spec2})
reduces to
the continuous eigenvalue
problem
\mbox{$\Psi_x=U\Psi$} with (\ref{spec2}) in a
suitable
continuous limit (cf.\ ref.~\citen{Ab78}).

It should be noted that the two systems (\ref{discrete1}) 
and (\ref{discrete3}) are connected through the change of variables 
\[
u_n^{(i)} \to (-1)^n u_n^{(i)}, \hspace{5mm} t \to -t. 
\]
This correspondence is not surprising 
as their respective 
``ancestor'' systems 
(\ref{sdmKdV}) and (\ref{sdKN}) 
are connected through the same type of transformation~\cite{Tsuchida4}. 

\subsection{Massive Thirring-like model 
$(\ref{cdMTM})$}

A space discretization of the matrix 
massive Thirring-type 
model 
(\ref{KMTM}), 
\begin{equation}
\left\{ \hspace{-1mm}
\begin{array}{l}
\QK_{n,\tau} + \i (\PHK_n+ \PHK_{n+1}) 
	+ 2 (\QK_n \CHK_{n+1} \PHK_n + \PHK_{n+1} \CHK_{n+1} \QK_n) = O,
\\[2pt]
\RK_{n,\tau} - \i (\CHK_n+ \CHK_{n+1}) 
	- 2 (\RK_n \PHK_{n} \CHK_n + \CHK_{n+1} \PHK_{n} \RK_n) = O,
\\[2pt]
\PHK_{n} - \PHK_{n+1} + \i \QK_n =O,
\\[2pt]
\CHK_{n} - \CHK_{n+1} - \i \RK_n =O,
\end{array}
\right.
\label{KNm}
\end{equation}
together with its Lax pair, 
was proposed in ref.~\citen{Tsuchida4}. 
%
%
This is considered 
the first 
negative flow of the semi-discrete Kaup--Newell 
hierarchy, which 
contains 
(\ref{sdKN}) as a positive flow. 
It should be noted 
that 
system (\ref{KNm}) permits 
the reduction 
\mbox{$\RK_n = C \QK_{n-\hf}^{\; T}$}, 
\mbox{$\CHK_n = -C \PHK_{n-\hf}^{\; T}$}, 
\mbox{$C^T = -C$}. 
In particular, the vector reduction 
\begin{align}
& \QK_n = (u_n^{(1)}, \ldots, u_n^{(M)}), \hspace{5mm}
\RK_n = C \QK_{n-\hf}^{\hspace{1pt}T},
\nn \\
& \PHK_n = \i (v^{(1)}_n, \ldots, v^{(M)}_n), \hspace{5mm}
\CHK_n = -C \PHK_{n-\hf}^{\; T}, \hspace{5mm} C^T = -C
\nn 
\end{align}
simplifies 
(\ref{KNm}) to a single vector equation,
\begin{align}
& \Ds \frac{\6}{\6 \tau}
 (v^{(i)}_n - v^{(i)}_{n+1}) + v^{(i)}_{n} + v^{(i)}_{n+1} 
 - 2 \biggl[ \sum_{1 \le j < k \le M} C_{jk} (v^{(j)}_{n+1} v^{(k)}_{n+\hf} 
 - v^{(j)}_{n+\hf} v^{(k)}_{n+1}) \biggr] v^{(i)}_{n+1}
\nn \\
& \hspace{10mm} 
\Ds \mbox{} - 2 \biggl[ 
  \sum_{1 \le j < k \le M} C_{jk} (v^{(j)}_{n+\hf} v^{(k)}_{n} 
 - v^{(j)}_{n} v^{(k)}_{n+\hf}) \biggr] v^{(i)}_{n} =0,
\hspace{5mm} i=1, 2,\ldots, M,
\label{}
\end{align}
with 
\mbox{$u^{(i)}_n 
= \boldsymbol{\Delta}_n v^{(i)}_{n} 
$}. 
This 
is a 
three-point 
($n$, $n+\hf$, $n+1$) 
difference scheme for 
the massive Thirring-like model (\ref{cdMTM}), 
and 
is considered 
a non-evolutionary potential 
symmetry of system (\ref{discrete2}). 
In the same 
manner as described in subsection \ref{type2}, 
we can 
also 
obtain 
a simpler, 
two-point difference scheme for 
(\ref{cdMTM}) from the 
matrix system (\ref{KNm}). 
Indeed, 
if we set 
\begin{equation}
\begin{split}
&
\QK_n = (u_n^{(1)}, \ldots, u_n^{(M)}), \hspace{5mm}
\RK_n = C (\QK_{n} + \QK_{n-1})^T, 
\\
& \PHK_n = \i (v^{(1)}_n, \ldots, v^{(M)}_n), \hspace{5mm} 
\CHK_n = - C (\PHK_n + \PHK_{n-1})^T, 
\hspace{5mm} C^T= -C, 
\label{qrpc}
\end{split}
\end{equation}
and 
utilize the relation 
\mbox{$\PHK_n C \PHK_m^T + \PHK_m C \PHK_n^T =0$}, 
(\ref{KNm}) is reduced to 
an alternative 
space 
discretization of 
(\ref{cdMTM}), 
\begin{align}
& \frac{\6}{\6 \tau} (v^{(i)}_{n} 
- v^{(i)}_{n+1}) + v^{(i)}_{n} + v^{(i)}_{n+1} 
 - 2 \biggl[ \sum_{1 \le j < k \le M} C_{jk} (v^{(j)}_{n+1} v^{(k)}_{n} 
 - v^{(j)}_{n} v^{(k)}_{n+1}) \biggr] (v^{(i)}_n + v^{(i)}_{n+1}) =0,
\nn \\
&  \hspace{110mm} i=1, 2,\ldots, M, 
\label{sdspMTM}
\end{align}
where 
\mbox{$u^{(i)}_n = 
\boldsymbol{\Delta}_n v^{(i)}_{n} 
(=v^{(i)}_{n+1} - v^{(i)}_{n})$}. 
This is 
considered 
a non-evolutionary symmetry of 
(the potential form of) system (\ref{discrete3}). 
The Lax pair for (\ref{sdspMTM}) is given by 
(cf.~(\ref{Lax_KN3}) and (\ref{LaxMTM})) 
\begin{subequations}
\begin{align}
L_n &= \left[
\begin{array}{cc}
 \mu & \left( \mu +1 \right) \left( \vt{v}_{n+1}-\vt{v}_n \right) 
\\[2pt]
 \left( -\mu + 1 \right) C 
\hspace{-1pt}\left( \vt{v}_{n+1}^T-\vt{v}_n^T \right) 
 & I - \left( \mu -1 \right)
C \hspace{-1pt}\left( \vt{v}_{n+1}^T-\vt{v}_n^T \right) 
 \left( \vt{v}_{n+1}-\vt{v}_n \right) 
\\
\end{array}
\right], 
\label{d-spec3}
\\[8pt]
 M_n &=  \frac{1}{\mu-1}
\left[
\begin{array}{cc}
 \mu +1
&
 -2 \left( \mu+1 \right) \vt{v}_n
\\
  -2 \left( \mu -1 \right) C\vt{v}_{n}^{T}
&
 4 \left( \mu -1 \right) C\vt{v}_{n}^{T} \vt{v}_{n}
\end{array}
\right],
\end{align}
\end{subequations}
where 
\mbox{$\vt{v}_n = (v_n^{(1)}, v_n^{(2)}, \ldots, v_n^{(M)})$}. 
Along parallel lines with 
the continuous case (cf.\ subsection~\ref{spMTM}), 
we can rewrite (\ref{sdspMTM}) as a closed 
differential-difference 
system for $u^{(i)}_n$. 
The resultant 
system 
provides an integrable semi-discretization of 
system (\ref{cdMTM3}). 

\subsection{Solutions to systems 
$(\ref{discrete3})$ and $(\ref{sdspMTM})$}
\label{discsoli}

In 
analogy with 
the continuous case (cf.\ (\ref{GLM_KN2})), 
a set of formulas for the solutions of the
space-discrete matrix 
Kaup--Newell 
system (\ref{sdKN})
as well as its commuting
flows, which 
tend to zero as \mbox{$n \to + \infty$}, can be presented 
in the following difference form~\cite{Tsuchi05}:
\begin{subequations}
\begin{align}
q_n 
&=
\boldsymbol{\Delta}_n {\cal K}(n,n
),\\[4pt]
r_n 
&=
\boldsymbol{\Delta}_n \bar{\cal K}(n,n
),\\[4pt]
{\cal K} (n,m) & =  
-\sum_{s=m}^{\infty
} \bar{F}(s) 
+ \sum_{j=0}^{\infty} \sum_{k=0}^{\infty}
\left[
{\cal K}(n,n+j) - {\cal K}(n,n+j+1)\right] F(n+j+k+1) \bar{F}(m+k)
\nn \\
&=
 \bar{G}(m)
+ \sum_{j=0}^{\infty} \sum_{k=0}^{\infty}
\left[ {\cal K}(n,n+j) - {\cal K}(n,n+j+1)\right]
\left[ G(n+j+k+1) - G(n+j+k+2) \right]
\nn \\
& \hphantom{=}
\hspace{25mm}
\mbox{}\times
\left[ \bar{G}(m+k) - \bar{G}(m+k+1) \right],
\hspace{5mm} m \geq n,
\\[4pt]
\bar{\cal K} (n,m)
&= 
- \sum_{s=m}^{\infty
} F(s) -
\sum_{j=0}^{\infty}
\sum_{k=0}^{\infty}
\left[ \bar{\cal K}(n,n+j) - \bar{\cal K}(n,n+j+1) 
\right] \bar{F}(n+j+k) F(m+k)
\nn \\
&=
 G(m) -
\sum_{j=0}^{\infty}
\sum_{k=0}^{\infty}
\left[ \bar{\cal K}(n,n+j) - \bar{\cal K}(n,n+j+1) \right]
\left[ \bar{G}(n+j+k) -\bar{G}(n+j+k+1)
\right]
\nn \\
& \hphantom{=}
\hspace{25mm}
\mbox{} \times
\left[ G(m+k) - G(m+k+1) \right],
\hspace{5mm} m \geq n.
\end{align}
\label{sdKNsol}
\end{subequations}
Here, the functions 
$\bar{F}(n)$ and $F(n)$ 
satisfy the corresponding {\it linear} \/uncoupled
system of matrix
differential-difference equations, 
{\it e.g.},
\begin{align}
\frac{\6 \bar{F}(n)}{\6 t} + \bar{F}(n+1) - \bar{F}(n-1)=O, 
\hspace{5mm}
\frac{\6 F(n)}{\6 t} + F(n+1) - F(n-1)=O
\label{3rdsdKN}
\end{align}
for the flow 
(\ref{sdKN}), 
and decay rapidly as \mbox{$n \to + \infty$}. 
The 
matrices $\bar{G}(n)$ and $G(n)$ are 
the 
``primitive functions'' of 
$\bar{F}(n)$ and $F(n)$, respectively, 
that also decay 
as \mbox{$n \to + \infty$} and 
satisfy the same linear system, 
that is, 
\mbox{$\bar{G} (n) :=-\sum_{s=n}^\infty \bar{F}(s)$} 
and \mbox{$G(n) :=-\sum_{s=n}^\infty F(s)$}. 
The reduction (\ref{QRn1}) 
is realized
at the level of the
solution formulas (\ref{sdKNsol}) 
by 
setting
\begin{align}
\bar{G} (n) = (g_1, g_2, \ldots, g_M)(n) =: \vt{g}(n),
\hspace{5mm} 
G (n) = 
C \bigl[ \bar{G}(n) + \bar{G}(n-1) \bigr]^T.
\label{FbarFKN}
\end{align}
In particular, the solutions to the
semi-discrete
coupled derivative mKdV equations (\ref{discrete3}), 
decaying as \mbox{$n \to + \infty$}, can be constructed
from those of the
linear vector differential-difference equation
\mbox{$\partial\vt{g}(n)/\partial t+\vt{g}(n+1) - \vt{g}(n-1)=\vt{0}$}
through the compact formula
\begin{subequations}
\begin{align}
\vt{u}_n(t)
&=
\boldsymbol{\Delta}_n \vt{k}(n, n;t),
\\[2pt]
\vt{k}(n, m)
&=
 \vt{g} (m) +
\sum_{j=0}^\infty \sum_{l=0}^\infty
\bigl[ \vt{k}(n, n+j) - \vt{k}(n, n+j+1) \bigr]
C \bigl[ \vt{g} (n+j+l) - \vt{g} (n+j+l+2)\bigr]^T
\nn \\
& \hphantom{=} 
\hspace{21mm}\mbox{}
\times
\bigl[ \vt{g} (m+l) - \vt{g} (m+l+1) \bigr], \hspace{5mm} m \ge n.
\label{}
\end{align}
\label{sdlinear2}
\end{subequations}
Here, \mbox{$\vt{u}_n = (u_n^{(1)}, u_n^{(2)}, \ldots, u_n^{(M)})$} and
\mbox{$\vt{k}(n,m)$}
are $M$-component row vectors. 
Substituting the expressions
\begin{align}
\vt{g} (n,t) &= \vt{a}_1
\mu_1^{-n} \e^{(\mu_1-\mu_1^{-1}) t}
+ \vt{a}_2 \mu_2^{-n} \e^{(\mu_2-\mu_2^{-1}) t},
\quad
|\mu_j|>1 \;(j=1,2),
\quad
\mu_1 \neq \mu_2,
\quad
\langle \vt{a}_1 C, \vt{a}_2 \rangle
\neq 0,
\nn \\
\vt{k}(n,m;t) &=
 \vt{k}_1 (n,t) \mu_1^{-m} \e^{(\mu_1-\mu_1^{-1}) t}
   + \vt{k}_2 (n,t) \mu_2^{-m} \e^{(\mu_2-\mu_2^{-1}) t}
\nn
\end{align}
into (\ref{sdlinear2}) and solving it
with respect to
$\vt{k}_1$ and $\vt{k}_2$,
we obtain the
``unrefined''
one-soliton solution of
system (\ref{discrete3}) in the difference 
form
\begin{align}
\vt{u}_n (t) = \boldsymbol{\Delta}_n 
\left\{
\frac{\vt{a}_1 \mu_1^{-n} \e^{ (\mu_1-\mu_1^{-1})t}
+\vt{a}_2 \mu_2^{-n} \e^{ (\mu_2 -\mu_2^{-1})t}}
{1- \frac{(\mu_1-\mu_2)(1-\mu_1)(1-\mu_2)}{(1-\mu_1 \mu_2)^2}
\sca{\vt{a}_1 C}{\vt{a}_2}
\mu_1^{-n} \mu_2^{-n} \e^{ (\mu_1 +\mu_2-\mu_1^{-1}-\mu_2^{-1})t}} 
\right\}.
\label{onesol7}
\end{align}
Note that the denominator in
the expression (\ref{onesol7})
may 
become zero for certain values of $n$ and $t$.
By introducing a new parametrization,
\begin{equation}
\begin{split}
& 
-\frac{(\mu_1-\mu_2)(1-\mu_1)(1-\mu_2)}{(1-\mu_1 \mu_2)^2}
\sca{\vt{a}_1 C}{\vt{a}_2}
=: \e^{-2 \delta} \;\,(\delta \in {\mathbb C}),
\\ 
& \vt{a}_1 =: 2 \e^{-\delta} \vt{b}_1, \hspace{5mm}
\vt{a}_2 =: 2 \e^{-\delta} \vt{b}_2,
\hspace{5mm}
\mu_1 = \e^{\alpha - \i \beta},
\hspace{5mm}
\mu_2 = \e^{\alpha + \i \beta},
\end{split}
\label{abparameter}
\end{equation}
(\ref{onesol7}) can be rewritten as
\begin{align}
\vt{u}_n (t)
&= 
\boldsymbol{\Delta}_n \left\{
\frac{\vt{b}_1 \e^{\i \beta n -2\i(\cosh \alpha \hspace{1pt}\sin \beta)
 t}
+\vt{b}_2 \e^{-\i \beta n + 2\i(\cosh \alpha \hspace{1pt}\sin \beta) t}
}
{\cosh  \left[
\alpha n -2 (\sinh \alpha \hspace{1pt}\cos \beta) t +\delta
 \right]} 
\right\} 
\nn \\[3pt]
&= 
\frac{
2\i 
\sqrt{\sin \left( \frac{\beta + \i \alpha}{2} \right)
\sin \left( \frac{\beta - \i \alpha}{2}\right)}}
{\cosh^2  \left[
\alpha n -2 (\sinh \alpha \hspace{1pt}\cos \beta) t + \delta' 
 \right] + \sinh^2\bigl( \frac{\alpha}{2} \bigr)
} 
\Bigl\{ \vt{b}'_1
\cosh  \left[
\alpha n -2 (\sinh \alpha \hspace{1pt}\cos \beta) t + \delta'
 + \i \varphi \right]
\nn \\ & 
\hphantom{=}
\mbox{} \times 
\e^{\i \beta n -2\i(\cosh \alpha \hspace{1pt}\sin \beta)t}
- \vt{b}'_2 
\cosh  \left[
\alpha n -2 (\sinh \alpha \hspace{1pt}\cos \beta) t + \delta'
 - \i \varphi \right]
\e^{-\i \beta n + 2\i(\cosh \alpha \hspace{1pt}\sin \beta) t}
\Bigr\},
\label{onesol8}
\end{align}
with the condition 
\mbox{$4\i (\cosh \alpha - \cos \beta)\sin \beta \hspace{2pt}
\sca{\vt{b}_1 C}{\vt{b}_2} = (\sinh \alpha)^2 $}. 
The constant $\varphi$ 
on the right-hand side of (\ref{onesol8}) 
is defined as 
\[
\exp(\i \varphi) := 
\frac{\sin \left( \frac{\beta + \i \alpha}{2} \right)}
{\sqrt{\sin \left( \frac{\beta + \i \alpha}{2} \right)
\sin \left( \frac{\beta - \i \alpha}{2}\right)}},
\]
and the 
new 
shifted parameters 
$\delta'$, 
$\vt{b}'_1$, and $\vt{b}'_2$ 
are given by \mbox{$\delta' 
:= \delta + \frac{\alpha}{2}$}, 
\mbox{$\vt{b}'_1 := \e^{\i \frac{\beta}{2}} \vt{b}_1$}, 
and \mbox{$\vt{b}'_2 := \e^{-\i \frac{\beta}{2}} \vt{b}_2$}. 
If we impose the ``reality conditions'' 
\mbox{$\alpha >0$},
\mbox{$0< \beta < \pi$} 
(or \mbox{$-\pi< \beta < 0$}), 
and
\mbox{$\e^{2\delta} \notin {\mathbb R}_{<0}$},
(\ref{onesol8}) provides 
the bright
one-soliton solution
of (\ref{discrete3}) 
that is indeed regular
for real $n$ and $t$.

A 
set of formulas for the solutions 
of the 
first 
negative flow (\ref{KNm}) 
of the semi-discrete matrix Kaup--Newell hierarchy, 
decaying as \mbox{$n \to + \infty$}, 
is completed by 
supplementing 
(\ref{sdKNsol}) 
with the 
following: 
\bseq
\begin{align}
- \i\hspace{1pt}\PHK_{n} 
&= 
{\cal K} (n,n),
\\
\i\hspace{1pt}\CHK_{n} 
&= 
\bar{\cal K} (n,n).
\end{align}
\label{appd}
\eseq
The 
{\it linear} \/uncoupled 
system of matrix differential-difference equations 
to be satisfied by
$\bar{F}(n)$ and $F(n)$ 
in 
this flow
reads as 
\begin{align}
\frac{\6 \bar{F}(n)}{\6 \tau} - \frac{\6 \bar{F}(n+1)}{\6 \tau}  
+ \bar{F}(n) + \bar{F}(n+1)=O, 
\hspace{5mm}
\frac{\6 F(n)}{\6 \tau} - \frac{\6 F(n+1)}{\6 \tau} 
+ F(n) + F(n+1)=O,
\nn 
\end{align}
and the same 
relation applies 
for their 
``primitive functions'' $\bar{G}(n)$ and $G(n)$.
Assuming the same 
restriction 
as that 
in the positive flow case (cf.\ (\ref{FbarFKN})), 
we can realize the reduction (\ref{qrpc}) 
on 
the solution 
formulas (\ref{sdKNsol}) and (\ref{appd}). 
Thus, 
the solutions to the
semi-discrete 
Thirring-like 
model (\ref{sdspMTM}), 
decaying as \mbox{$n \to + \infty$}, can be constructed
from those of the
linear vector differential-difference equation
\mbox{$\partial\vt{g}(n)/\partial \tau 
- \partial\vt{g}(n+1)/\partial \tau 
+\vt{g}(n) + \vt{g}(n+1)=\vt{0}$}
through the compact formula
\begin{subequations}
\begin{align}
\vt{v}_n(\tau)
&=
\vt{k}(n, n;\tau),
\\[2pt]
\vt{k}(n, m)
&=
 \vt{g} (m) +
\sum_{j=0}^\infty \sum_{l=0}^\infty
\bigl[ \vt{k}(n, n+j) - \vt{k}(n, n+j+1) \bigr]
C \bigl[ \vt{g} (n+j+l) - \vt{g} (n+j+l+2)\bigr]^T
\nn \\
& {\hphantom =} \hspace{21mm}\mbox{}
\times
\bigl[ \vt{g} (m+l) - \vt{g} (m+l+1) \bigr], \hspace{5mm} m \ge n. 
\label{}
\end{align}
\label{sdlinear3}
\end{subequations}
Here, 
\mbox{$\vt{v}_n = (v_n^{(1)}, v_n^{(2)}, \ldots, v_n^{(M)})$}. 
Using the formula (\ref{sdlinear3}), we can construct the 
soliton solutions
of system (\ref{sdspMTM}) in 
a manner similar to that 
in the positive flow case.
In particular, 
the ``unrefined'' one-soliton 
solution of 
(\ref{sdspMTM}) is given by 
\[
\vt{v}_n (\tau)
= 
\frac{\vt{a}_1 \mu_1^{-n} \e^{-\frac{\mu_1+1}{\mu_1-1} \tau}
+\vt{a}_2 \mu_2^{-n} \e^{-\frac{\mu_2+1}{\mu_2-1} \tau}}
{1- \frac{(\mu_1-\mu_2)(1-\mu_1)(1-\mu_2)}{(1-\mu_1 \mu_2)^2}
\sca{\vt{a}_1 C}{\vt{a}_2}
\mu_1^{-n} \mu_2^{-n} \e^{-\bigl( \frac{\mu_1+1}{\mu_1-1} +
\frac{\mu_2+1}{\mu_2-1} \bigr) \tau}},
\]
which can be rewritten in terms of the parametrization (\ref{abparameter}) as 
\[
\vt{v}_n (\tau)
=
\frac{\vt{b}_1 \e^{\i \beta n -\i 
\frac{\scriptstyle\sin \beta}{2\sinh \left(\frac{\alpha+ \i \beta}{2} \right)
\sinh \left(\frac{\alpha- \i \beta}{2} \right) }\tau}
+\vt{b}_2 \e^{-\i \beta n + \i 
\frac{\scriptstyle
\sin \beta}{2\sinh \left(\frac{\alpha+ \i \beta}{2} \right)
\sinh \left(\frac{\alpha- \i \beta}{2} \right) }\tau}}
{\cosh  \left[
\alpha n 
+\frac{
\sinh \alpha}{2\sinh \left(\frac{\alpha+ \i \beta}{2} \right)
\sinh \left(\frac{\alpha- \i \beta}{2} \right) } \tau +\delta
 \right]}, 
\]
with the condition 
\mbox{$4\i (\cosh \alpha - \cos \beta)\sin \beta \hspace{2pt}
\sca{\vt{b}_1 C}{\vt{b}_2} = (\sinh \alpha)^2 $}. 
This provides 
the 
bright 
one-soliton solution 
under the 
``reality conditions'' 
\mbox{$\alpha >0$},
\mbox{$0< \beta < \pi$} (or \mbox{$-\pi< \beta < 0$}),  
and
\mbox{$\e^{2\delta} \notin {\mathbb R}_{<0}$}, 
which is indeed 
regular for real $n$ and $\tau$. 

\section{Concluding remarks}
\label{}
\setcounter{equation}{0}

In this paper, we have 
proposed 
a new type of reduction 
involving an antisymmetric constant matrix; this reduction 
relates one matrix 
variable with another in 
a 
system of 
coupled matrix PDEs. 
In the particular 
case of vector variables, 
it enables us 
to obtain the 
integrable vector 
PDEs having $Sp(m)$ as 
their 
symmetry group. 
Our approach 
has been proven 
to be applicable to 
both continuous and 
discrete systems. 
For the 
particularly 
interesting systems 
such as 
(\ref{cdmkdv1}), (\ref{cdmkdv2}), 
(\ref{cdMTM}), (\ref{discrete1}), 
(\ref{discrete3}), and (\ref{sdspMTM}), the 
one-soliton solutions are derived 
from 
the (discrete) 
linear integral equations of the Gel'fand--Levitan--Marchenko type. 
The solutions 
are clearly 
invariant 
up to a redefinition of the 
soliton parameters 
under the same symmetry group 
$Sp(m)$ as that of 
the original systems. 
The behavior of the solitons 
reflects 
an interesting 
characteristic
of these systems; 
the 
total ``particle number'' of the system, 
{\it e.g.},
$\int_{-\infty}^{\infty} 
\| \vt{u} \|^2\hspace{1pt}
\d x$ in the continuous case, is, in general, 
not conserved 
and varies in time. 
As a result, each soliton exhibits 
an overall vectorial 
oscillation, 
in addition to 
an internal oscillation among the vector components. 
A 
detailed investigation of the multi-soliton solutions 
would be an 
interesting and promising path 
toward 
the construction of a new class of 
set-theoretical solutions with symplectic invariance 
to the quantum Yang--Baxter 
equation (cf.\ ref.~\citen{Tsuchi04}). 

One natural 
question 
arising from 
the results 
of 
subsections~\ref{CLLtype} and \ref{KNtype}
concerns the integrability of a 
system of 
the following general form: 
\begin{align}
\frac{\6 u_{i}}{\6 t} + \frac{\6^3 u_{i}}{\6 x^3}
& + a \Biggl[ \sum_{1 \le j<k \le M}
        C_{jk} \Bigl( \frac{\6 u_{j}}{\6 x} u_{k} - u_{j}
        \frac{\6 u_{k}}{\6 x} \Bigr) \Biggr] \frac{\6 u_{i}}{\6 x} 
\nn \\
& \mbox{} + b \frac{\6}{\6 x} \Biggl[
        \sum_{1 \le j<k \le M} C_{jk} \Bigl( \frac{\6 u_{j}}{\6 x} u_{k}
        - u_{j} \frac{\6 u_{k}}{\6 x} \Bigr) u_{i} \Biggr] = 0,
\hspace{5mm}
i=1, 2, \ldots, M,
\label{cdmkdv3}
\end{align}
where $a$ and $b$ are constants. 
Using the 
{\it Mathematica} \/package
``InvariantsSymmetries.m''~\cite{Hereman}, 
we 
searched for 
the 
cases wherein 
system (\ref{cdmkdv3}) 
with \mbox{$M=2$} 
can possess
higher 
polynomial 
conservation laws 
and/or 
symmetries 
of the 
prescribed orders. 
The result was null in that 
within the limitations of 
our computer's memory and CPU performance, 
we could 
detect no integrable case except 
the 
already 
found two cases \mbox{$a=0$} and \mbox{$b=0$}. 
Our future task is 
to 
rigorously 
prove (or disprove) 
the 
nonintegrability of 
system (\ref{cdmkdv3}) 
in the case 
\mbox{$ab \neq 0$}. 

We 
would like to 
explain 
how 
the 
semi-discrete systems 
of the form 
$\6_t u^{(i)}_{n} + (u^{(i)}_{n+1} - 
u^{(i)}_{n-1}) + \big\{\mbox{nonlinear terms}\bigr\}=0$
presented in section~\ref{IntDisc} 
can be 
related to 
the third-order PDEs 
of the form $\6_T u_{i} + \6_X^{\hspace{1pt}3} u_{i} +
\big\{\mbox{nonlinear terms}\bigr\}=0$ 
in a 
continuous limit. 
In fact, 
the asymptotic 
expansion 
with respect to the space interval 
$\varDelta$, 
$\, \6_t u^{(i)}_{n} + (u^{(i)}_{n+1} - u^{(i)}_{n-1}) 
\simeq 
\6_t u^{(i)} + 2 \varDelta \6_x u^{(i)} 
	+ \frac{1}{3} \varDelta^3 \6_x^3 u^{(i)} + O(\varDelta^5)$, 
implies that 
a 
Galilean 
plus 
scaling transformation 
such as \mbox{$\6_t + 2\varDelta \6_x =: \frac{1}{3} \varDelta^3 \6_T$}, 
\mbox{$\6_x =: \6_X$}, 
\mbox{$u^{(i)}
\hspace{-1pt} 
\sim 
\hspace{-1pt}
O(\varDelta^\hf)$},
or equivalently, 
\mbox{$u_n^{(i)} (t)
\hspace{-1pt} 
\sim 
\hspace{-1pt} 
\varDelta^\hf u_i 
	\bigl(\varDelta (n-2t), 
\varDelta^3 t/3 \bigr)$} 
must 
be performed. 
This 
is a commonly 
accepted technique 
(see, {\em e.g.}, refs.~\citen{Ab78,AL76}), 
and the 
solutions 
of 
such a semi-discrete system 
generally 
have 
the same structures as 
those of the 
corresponding 
continuous 
system. 
However, one can also obtain 
further 
``natural'' 
space discretizations 
that directly arrive at 
the third-order PDEs 
in the continuum limit, 
without resorting to 
the Galilean transformations. 
This is achieved 
by considering 
a 
proper 
linear combination of 
the original 
semi-discrete system 
and a 
higher symmetry of 
it~\cite{AL76,Sur97}. 
Note that 
this ``improvement'' 
results 
only in a minor change 
in the time dependence of certain 
parameters 
in the 
solutions, 
while 
a 
semi-discrete system 
obtained in this manner 
usually 
appears rather complicated 
and less attractive than the original one. 
Therefore, 
we do not pursue such a direction in this paper. 

We have concentrated on the 
reductions of the third-order (\mbox{$\omega \propto k^3$}) 
flows 
of the 
derivative NLS (DNLS)-type hierarchies as well as their 
first negative 
(\mbox{$\omega \propto k^{-1}$})
flows, 
in both 
the continuous 
and 
semi-discrete cases. 
The feasibility of the 
reduction is based on the 
fact that 
the cubic terms 
in the evolution 
equation for $\QK$ 
and those 
for $\RK$ 
in such systems, 
{\it e.g.}, (\ref{higherCLL}) or 
(\ref{higherKN}), 
have opposite signs. 
This is in contrast with 
the case of 
the corresponding 
flows of 
the matrix 
NLS hierarchy, 
{\it e.g.}, the 
(non-reduced) matrix complex mKdV equation (\ref{cmkdv})
that permits 
various reductions, 
including 
\mbox{$R = A_1 Q^T A_2$} with 
\mbox{$A_1^T=A_1$} and \mbox{$A_2^T=A_2$} 
or \mbox{$A_1^T=-A_1$} and \mbox{$A_2^T=-A_2$}, 
but not the reduction \mbox{$R = C Q^T$} or \mbox{$R = Q^T C$} 
with \mbox{$C^T=-C$}. 
However, it should be noted 
that the matrix DNLS flows 
are not the only class 
of systems that 
permit the 
reductions 
of the latter type 
exploited 
in this paper. 
As an illustrative 
example, 
let us consider 
the matrix generalization
of the Yajima--Oikawa system~\cite{YO} 
(cf.\ refs.~\citen{YCMa,Melnikov,Kaup,Strampp,Cheng,Liu}), 
\begin{equation}
\left\{
\begin{array}{l}
 \i Q_{t_2} + Q_{xx} - PQ =O, \\
 \i R_{t_2} - R_{xx} + RP =O, \\
 \i P_{t_2} + 2 (QR)_x = O, 
\end{array}
\right.
\label{mYO}
\end{equation}
and its third-order symmetry. 
System (\ref{mYO}) possesses 
a Lax pair of the form 
\bseq
\begin{align}
U &=
\i \z \left[
\begin{array}{ccc}
 -I_1 & & \\
  & O & \\
 & & I_1 \\
\end{array}
\right] + 
\left[
\begin{array}{ccc}
 O & Q & P \\
 O & O & R \\
 I_1 & O & O \\
\end{array}
\right],
\label{LmYO}
\\[8pt]
V &=
\i \z^2 
\left[
\begin{array}{ccc}
 O & & \\
 & I_2 & \\
 & & O \\
\end{array}
\right] + \z 
\left[
\begin{array}{ccc}
 O & Q & O \\
 O & O & -R \\ 
 O & O & O \\ 
\end{array}
\right] + 
\i \left[
\begin{array}{ccc}
O & Q_x & QR \\
R & O & -R_x \\
O & Q & O \\
\end{array}
\right].
\end{align}
\label{LYO}
\eseq
This implies 
that the 
substitution of 
(\ref{LYO}) 
in 
the zero-curvature condition (\ref{Lax_eq}) 
yields (\ref{mYO}). 
The matrix Yajima--Oikawa system (\ref{mYO}) 
allows 
the standard 
reduction \mbox{$R = 
B Q^\dagger$}, \mbox{$P^\dagger=P$}, 
\mbox{$B^\dagger = -B$}, \mbox{$B_t=B_x=O$}. 
The third-order symmetry of 
(\ref{mYO}) reads as 
\begin{equation}
\left\{
\begin{array}{l}
 Q_{t_3} + 4 Q_{xxx} - 3P_x Q - 6PQ_x -6QRQ =O, \\
 R_{t_3} + 4 R_{xxx} - 3RP_x -6 R_x P + 6RQR =O, \\
 P_{t_3} + P_{xxx} -3 (P^2)_x + 6 (Q_{x} R -QR_{x})_x = O, 
\end{array}
\right.
\label{mYO2}
\end{equation}
and the corresponding Lax pair 
is given by 
(\ref{LmYO}) and 
\begin{align}
V &=
\i \z^3 
\left[
\begin{array}{ccc}
 -4I_1 & & \\
 & O & \\
 & & 4I_1 \\
\end{array}
\right] + \z^2
\left[
\begin{array}{ccc}
 O & 4Q & 4P \\
 O & O & 4R \\ 
 4I_1 & O & O \\ 
\end{array}
\right] + \i \z
\left[
\begin{array}{ccc}
-2P & 4 Q_x & 2P_x \\
 O & O & 4R_x \\
 O & O & 2P \\
\end{array}
\right] 
\nn \\
& {\hphantom =} \mbox{}+
\left[
\begin{array}{ccc}
 P_x + 2QR & -4Q_{xx} + 2PQ & -P_{xx} + 2P^2 + 2(QR_x - Q_x R) \\
 4R_x & -4 RQ & -4 R_{xx} + 2RP \\
 2P & -4Q_x & -P_x + 2QR \\
\end{array}
\right].
\nn
\end{align}
System (\ref{mYO2}) 
allows 
an extension of 
the typical reduction 
considered 
in this paper, that is, 
\mbox{$R=CQ^T$} and 
\mbox{$P^T =P$} or \mbox{$R= Q^T C$} and \mbox{$P^T C=CP$}, 
where $C$ is an antisymmetric constant matrix. 
In particular, the 
vector reduction in the former case 
changes the matrix system (\ref{mYO2}) into 
the following system~\cite{Melnikov,Loris2}:
\begin{equation}
\left\{
\begin{array}{l}
 \Ds \frac{\6 u_i}{\6 t} + 4 \frac{\6^3 u_{i}}{\6 x^3} 
 - 3 \frac{\6 p}{\6 x} u_i -  6 p \frac{\6 u_{i}}{\6 x} =0, \hspace{5mm} 
  	i=1, 2, \ldots, M,\\[10pt]
 \Ds \frac{\6 p}{\6 t} + \frac{\6^3 p}{\6 x^3}
 -6 p \frac{\6 p}{\6 x} + 12 \sum_{1 \le j < k \le M}
	C_{jk} \Bigl( \frac{\6^2 u_{j}}{\6 x^2} u_k - u_j 
	\frac{\6^2 u_{k}}{\6 x^2} \Bigr) = 0.
\end{array}
\right.
\nn
\end{equation}
It is noted that
this system 
is a modification 
of the triangular 
system comprising 
the KdV equation and 
time part of the associated linear problem
due to the 
addition of the last summation term. 

\section*{Acknowledgments}
The author is grateful to Dr.\ Ken-ichi Maruno, 
Dr.\ Zengo Tsuboi, 
and 
Dr.\ Yukitaka Ishimoto for 
their 
useful comments. 

\addcontentsline{toc}{section}{References} 


\begin{thebibliography}{99}

\bibitem{Manakov}
S.~V.~Manakov: 
Zh.\ Eksp.\ Teor.\ Fiz.\ {\bf 65} (1973) 505
[Translation:\ Sov.\ Phys.\
JETP {\bf 38} (1974) 248].

\bibitem{ZaSha}
V.\ E.\ Zakharov and A.\ B.\ Shabat: 
Funct.\ Anal.\ Appl.\ {\bf 8} (1974) 226. 

\bibitem{Svip}
S.\ I.\ Svinolupov: Funct.\ Anal.\ Appl.\ {\bf 27} (1993) 257. 

\bibitem{YO2}
N.\ Yajima and M.\ Oikawa: Prog.\ Theor.\ Phys.\ {\bf 54} (1975) 1576.

\bibitem{Konop3}
B.\ G.\ Konopelchenko: Phys.\ Lett.\ B {\bf 100} (1981) 254. 

\bibitem{Adler}
V.\ E.\ Adler: 
Physica D {\bf 87} (1995) 52. 

\bibitem{Tsuchida5}
T.\ Tsuchida and M.\ Wadati: J.\ Phys.\ Soc.\ Jpn.\ {\bf 68} (1999) 2241.

\bibitem{WKI}
M.\ Wadati, K.\ Konno and Y.\ H.\ Ichikawa: 
J.\ Phys.\ Soc. Jpn.\ {\bf 47} (1979) 1698. 


\bibitem{SW}
V.\ V.\ Sokolov and T.\ Wolf: J.\ Phys.\ A:\ Math.\ Gen.\ {\bf 34} 
(2001) 11139.

\bibitem{TsuWo}
T.\ Tsuchida and T.\ Wolf: J.\ Phys.\ A:\ Math.\ Gen.\ {\bf 38} 
(2005) 7691.

\bibitem{AnWo}
S.\ Anco and T.\ Wolf: J.\ Nonl.\ Math.\ Phys.\ 
{\bf 12} Suppl.\ 1 (2005) 13.

\bibitem{Iwao2} M. Iwao and R. Hirota: J. Phys. Soc. Jpn. {\bf 69} 
(2000) 59.

\bibitem{Loris1}
I.\ Loris and R.\ Willox: J.\ Math.\ Phys.\ {\bf 40} (1999) 1420.

\bibitem{Loris3}
I.\ Loris: {\it Proc.\ of the workshop on nonlinearity, 
integrability and all that:\ 
Twenty years after NEEDS '79} 
(World Scientific, Singapore, 2000) p.\ 325.

\bibitem{Gakkai}
T.\ Tsuchida: 
A talk presented at 
the 2002 Autumn 
Meeting of the Physical Society of Japan; 
recorded in 
Meeting abstracts of the Physical Society of Japan 
(2002). 

\bibitem{Nij1}
J.\ van der Linden, H.\ W.\ Capel and F.\ W.\ Nijhoff: Physica A 
{\bf 160} (1989) 235. 

\bibitem{Tsuchida3}
T.\ Tsuchida and M.\ Wadati: Inverse Probl.\ {\bf 15} (1999) 1363.

\bibitem{Kuz}
E.\ A.\ Kuznetsov and A.\ V.\ Mikhailov: 
Theor.\ Math.\ Phys.\ {\bf 30} (1977) 193. 

\bibitem{KN2}
D.\ J.\ Kaup and A.\ C.\ Newell: 
Lett.\ Nuovo Cimento {\bf 20} (1977) 325. 

\bibitem{KaMoIno}
T.\ Kawata, T.\ Morishima and H.\ Inoue: 
J.\ Phys.\ Soc.\ Jpn.\ {\bf 47} (1979) 1327.

\bibitem{GIK}
V.\ S.\ Gerdjikov, M.\ I.\ Ivanov and P.\ P.\ Kulish: Theor.\ Math.\ Phys.\ 
{\bf 44} (1980) 784. 

\bibitem{NCQL}
F.\ W.\ Nijhoff, H.\ W.\ Capel, G.\ R.\ W.\ Quispel and J.\ van der Linden: 
Phys.\ Lett.\ A {\bf 93} (1983) 455.

\bibitem{Tsuchida4}
T.\ Tsuchida: J.\ Phys.\ A:\ Math.\ Gen.\ {\bf 35} (2002) 7827.

\bibitem{Satake}
I.\ Satake: {\em Linear algebra}, 
translated by S.\ Koh, 
T.\ Akiba and S.\ Ihara (Marcel Dekker, New York, 1975). 

\bibitem{Serre}
D.\ Serre: {\em Matrices:\ theory and applications}, 
Graduate Texts in Mathematics Vol.\ 216 
(Springer, New York, 2002).

\bibitem{Tsuchi05}
T.\ Tsuchida: 
{\em Refinements of the Inverse Scattering Method} 
(a tentative title), in preparation. 

\bibitem{ARS}
M.\ J.\ Ablowitz, A.\ Ramani and H.\ Segur: 
J.\ Math.\ Phys.\ {\bf 21} (1980) 1006. 

\bibitem{CLL}
H.\ H.\ Chen, Y.\ C.\ Lee and C.\ S.\ Liu: 
Phys.\ Scr.\ {\bf 20} (1979) 490. 

\bibitem{Olver}
P.\ J.\ Olver and V.\ V.\ Sokolov: 
Commun.\ Math.\ Phys.\ {\bf 193} (1998) 245.

\bibitem{Dimakis}
A.\ Dimakis and F.\ M\"{u}ller-Hoissen: 
J.\ Phys.\ A:\ Math.\ Gen.\ {\bf 39} (2006) 14015.

\bibitem{AKNS74}
M.\ J.\ Ablowitz, D.\ J.\ Kaup, A.\ C.\ Newell and H.\ Segur:
Stud.\ Appl.\ Math.\ {\bf 53} (1974) 249.

\bibitem{ZS79}
V.\ E.\ Zakharov and A.\ B.\ Shabat:
Funct.\ Anal.\ Appl.\ {\bf 13} (1979) 166.


\bibitem{Makhankov}
V.\ G.\ Makhan'kov and O.\ K.\ Pashaev: 
Theor.\ Math.\ Phys. {\bf 53} (1982) 979.

\bibitem{IKWS}
Y.\ H.\ Ichikawa, K.\ Konno, M.\ Wadati and H.\ Sanuki: 
J.\ Phys.\ Soc.\ Jpn.\ {\bf 48} (1980) 279.

\bibitem{Kawata2}
T.\ Kawata, J.\ Sakai and N.\ Kobayashi: 
J.\ Phys.\ Soc.\ Jpn.\ {\bf 48} (1980) 1371.

\bibitem{Zakh}
V.\ E.\ Zakharov: 
{\em Solitons} \/ed.\ R.\ K.\ Bullough and P.\ J.\ Caudrey 
(Topics in Current Physics
17,
Springer, Berlin, 1980) p.\ 243.

\bibitem{Konop1}
B.\ G.\ Konopelchenko: Phys.\ Lett.\ A {\bf 79} (1980) 39. 

\bibitem{Tsuchida1}
T.\ Tsuchida and M.\ Wadati: J.\ Phys.\ Soc.\ Jpn.\ {\bf 67} (1998) 1175.

\bibitem{Kawa}
T.\ Kawata, N.\ Kobayashi and H.\ Inoue: J.\ Phys.\ Soc.\ Jpn.\ 
{\bf 46} (1979) 1008.

\bibitem{KN}
D.\ J.\ Kaup and A.\ C.\ Newell: J.\ Math.\ Phys.\ {\bf 19} (1978) 798. 

\bibitem{Konop4}
B.\ G.\ Konopelchenko: J.\ Phys.\ A:\ Math.\ Gen.\ {\bf 14} (1981) 3125. 

\bibitem{Kako}
F.\ Kako and N. Mugibayashi: Prog.\ Theor.\ Phys.\ {\bf 61} (1979) 776.

\bibitem{Ab78}
M.\ J.\ Ablowitz: Stud.\ Appl.\ Math.\ {\bf 58} (1978) 17.

\bibitem{GIV}
V.\ S.\ Gerdjikov, M.\ I.\ Ivanov and Y.\ S.\ Vaklev: 
Inverse Probl.\ {\bf 2} (1986) 413.

\bibitem{Tsuchi04}
T.\ Tsuchida: Prog.\ Theor.\ Phys.\ {\bf 111} (2004) 151.

\bibitem{Hereman}
\"{U}.\ G\"{o}kta\c{s} 
and W.\ Hereman: 
Adv.\ Comput.\ Math.\ {\bf 11} (1999) 55. 

\bibitem{AL76}
M.\ J.\ Ablowitz and J.\ F.\ Ladik: 
J.\ Math.\ Phys.\ {\bf 17} (1976) 1011. 
\bibitem{Sur97}
Y.\ B.\ Suris: 
Phys.\ Lett.\ A {\bf 234} (1997) 91. 

\bibitem{YO}
N.\ Yajima and M.\ Oikawa: Prog.\ Theor.\ Phys.\ {\bf 56} (1976) 1719.

\bibitem{YCMa}
Y.-C.\ Ma: Wave Motion {\bf 3} (1981) 257. 

\bibitem{Melnikov}
V.\ K.\ Mel'nikov: Lett.\ Math.\ Phys.\ {\bf 7} (1983) 129.

\bibitem{Kaup}
D.\ J.\ Kaup: 
{\it Nonlinear evolutions:\ 
Proc.\ of the IVth Workshop on Nonlinear 
Evolution Equations and Dynamical Systems} 
(Balaruc-les-Bains, 
June 1987) ed.\ Jerome J.\ P.\ Leon 
(World Scientific, Singapore, 1988) p.\ 161. 

\bibitem{Strampp}
J.\ Sidorenko and W.\ Strampp: J.\ Math.\ Phys.\ {\bf 34} (1993) 1429. 

\bibitem{Cheng}
Y.\ J.\ Zhang and Y.\ Cheng: J.\ Math.\ Phys.\ {\bf 35} (1994) 5869. 

\bibitem{Liu}
Q.\ P.\ Liu: hep-th/9502076. 

\bibitem{Loris2}
I.\ Loris: J.\ Phys.\ Soc.\ Jpn.\ {\bf 70} (2001) 662.

\end{thebibliography}
\end{document}